\documentclass[%
 reprint,
 amsmath,amssymb,
 aps,
 prb
]{revtex4-1}

\usepackage{graphicx}
\usepackage{dcolumn}
\usepackage{bm}
\usepackage[colorlinks]{hyperref}
\usepackage[caption=false]{subfig}
\usepackage[final]{changes} 

\newcommand{\fixme}[1]%
   {\begingroup{\color{blue}[NOTE: \textit{#1}]}\endgroup}

\begin{document}

\preprint{APS/123-QED}


\title{System--Environment Correlations in Qubit Initialization and Control}

\author{Jani Tuorila$^{1,2}$}
\author{J\"urgen Stockburger$^3$}
\author{Tapio Ala-Nissila$^{1,4,5}$}
\author{Joachim Ankerhold$^3$}
\author{Mikko M\"ott\"onen$^{1,6}$}

\affiliation{$^1$QCD Labs and MSP Group, QTF Centre of Excellence, Department of Applied Physics, Aalto University, P.O. Box 15100, FI-00076 Aalto, Finland\\
$^2$Nano and Molecular Materials Research Unit, University of Oulu, P.O. Box 3000, FI-90014 Oulu, Finland\\
$^3$Institute for Complex Quantum Systems and IQST, University of Ulm, D-89069 Ulm, Germany\\
$^4$Department of Physics, Box 1843, Brown University, Box 1843, Providence, Rhode Island 02912-1843, USA\\
$^5$Interdisciplinary Centre for Mathematical Modelling, Department of Mathematical Sciences, Loughborough University, Loughborough LE11 3TU, United Kingdom\\
$^6$VTT Technical Research Centre of Finland, QTF Centre of Excellence, P.O. Box 1000, FI-02044, Aalto, Finland}

\date{\today}

\begin{abstract}
\added{The impressive progress in fabricating and controlling superconducting devices for quantum information processing has reached a level where reliable theoretical predictions need to account for quantum correlations that are not captured by the conventional modeling of contemporary quantum computers. This applies particularly to the qubit initialization as the process which crucially limits typical operation times. Here we employ numerically exact methods to study realistic implementations of a transmon qubit embedded in electromagnetic environments focusing on the most important system-reservoir correlation effects such as the Lamb shift and entanglement. For the qubit initialization we find a fundamental trade-off between speed and accuracy which sets intrinsic constraints in the optimization of future reset protocols. Instead, the fidelities of quantum logic gates can be sufficiently accurately predicted by standard treatments. Our results can be used to accurately predict the performance of specific set-ups and also to guide future experiments in probing low-temperature properties of qubit reservoirs.}
\deleted{We employ numerically exact methods to study a qubit coupled to a quantum-mechanical bath, focusing on important correlation effects such as the Lamb shift and entanglement. \deleted{that are typically neglected in the modeling of contemporary quantum computers.} \added{In addition to an ideal qubit, we study a more general superconducting transmon qubit with $N$ levels.} We find a fundamental trade-off between speed and accuracy in qubit initialization which is important in the optimization of future reset protocols. Namely, we show analytically that at low temperatures, the deviation from the bare qubit ground state is proportional to the qubit decay rate, caused by the unavoidable entanglement between the qubit and the bath. We also find that the qubit decay is superexponential at short time scales, contrary to the result from standard Markovian approaches. However, the fidelities of quantum logic gates can be sufficiently accurately predicted by the Markovian methods. Our results can be used to develop quantum devices with engineered environments and to guide future experiments in probing the properties of the reservoirs.}
\end{abstract}

\pacs{Valid PACS appear here}
\maketitle


\section{Introduction}
Precise control and preparation of pure quantum states are pivotal in 
quantum technological applications of practical interest~\cite{divincenzo2000,nielsen2000}. 
For example, fast and high-fidelity initialization of a qubit to its ground state is required to realize a large-scale gate-based quantum computer since implementations of quantum error correction codes~\cite{shor1995,Fowler_2009,Fowler_2011} call for pure ancillary qubits at each error correction cycle. However, satisfactory qubit reset still remains a technological challenge. 
In the most promising approaches, the qubit is steered towards the desired state by coherent driving~\cite{valenzuela2006,grajcar2008,geerlings2013,jin2015} or by using a specifically tailored dissipative environment~\cite{jones2013,tuorila2017}. The latter has the benefit that its theoretical modeling does not rely on rotating frames which are often used in the case of time-dependent driving and can cause inaccuracies in the predicted figures of merit.

Fast initialization inherently calls for relatively strong environmental coupling, whereas coherent operations such as quantum logic are error free only in the limit of isolated quantum systems. This apparent conflict can be resolved with a dissipative low-temperature environment and temporal control over the coupling strength~\cite{jones2013,tuorila2017,Silveri}, providing very weak coupling \replaced{during}{for} the coherent control and strong coupling \replaced{during}{for} fast initialization to the ground state. 
Recently, superconducting-circuit realizations of the quantum-circuit refrigerator~\cite{tan2017,Silveri2018} and the tunable heat sink~\cite{partanen2018,partanen2018b} have demonstrated that with the current technology, one can indeed control the coupling strength between the quantum system and the engineered bath over several orders of magnitude with a minimal effect on the system frequency. Such components can be conveniently integrated on the same chip with qubits, allowing scalable fabrication and low circuit complexity.

Estimates of the speed and the fidelity of the initialization protocols based on qubit decay have been made in the weak-coupling, i.e., Born--Markov, approximation~\cite{jones2013,tuorila2017,geerlings2013}. Stationary states then arise from a detailed balance condition of the Born--Markov rates and, therefore, appear independent of the coupling strength. However, experiments with engineered quantum systems are entering a regime of high accuracy~\cite{Barends_2013,Kelly_2015}, where higher-order corrections need to be included. One such correction is the modification of equilibrium populations through a Lamb shift~\cite{fragner2008}, which can be sizable in the case of a broadband environment~\cite{Silveri2018,gramich2011,gramich2014}. System--environment entanglement is another higher-order effect detrimental to the performance of reset protocols. Any realistic analysis and optimization of the speed and fidelity of the envisioned protocols thus calls for an exact analysis of dissipation that goes beyond the conventional weak-coupling formalism. 

Here, we examine \added{non-perturbatively the} open quantum dynamics of \replaced{both an ideal two-level quantum system and a superconducting transmon qubit with $N$ energy levels~\cite{koch2007}.} {a qubit non-perturbatively and} 
\added{We focus on the figures of merit important to the quantum information community and leave, e.g., 
more detailed studies of non-Markovianity~\cite{breuer2016, rivas2014, wolf2008, luo2012, chruscinski2014, basilewitsch2017} for future work. We} demonstrate that the effects of entanglement and Lamb shift lead to a decreasing ground-state occupation in the steady state with increasing bath coupling. This indicates a potential need to make a compromise between speed and accuracy in qubit initialization protocols. We observe a further departure from the behavior predicted by Born--Markov master equations in the transient dynamics of an initially decoupled qubit, displayed as a rapid initial decoherence into a mixture of pointer states~\cite{zurek2003,braun2001}. For moderate and strong coupling, the initial transient dynamics has a Gaussian\replaced{ temporal shape}{-shaped time dependence} which is independent of the qubit frequency, also referred to as universal decoherence. In addition, we find qubit-reservoir entanglement to be the dominant source of initialization error at low temperatures, \replaced{whereas}{while} strong-coupling effects are minor for quantum gates at experimentally relevant parameter values. \added{During initialization using strong coupling to an engineered bath, effects of the intrinsic qubit dissipation are small and, thus, can be neglected here\replaced{ (see Appendix~\ref{app:intr})}{~\cite{SI}}.} Our findings can be used to improve qubit schemes involving reservoir engineering. 

This paper is organized as follows. In Sec.~\ref{sec:system}, we introduce a prototype system for studies of strong bath-coupling effects, consisting of a superconducting transmon qubit bilinearly coupled to a thermal bath. We also describe the numerically exact method used in our simulations. In Sec.~\ref{sec:relaxation}, we study the decay dynamics of the qubit and give a detailed description of the shortcomings of the Born--Markov master equations in terms of universal decoherence. Section~\ref{sec:steady} presents an accurate calculation of the steady state. We compare the numerically exact data against the Boltzmann distribution of the bare qubit, and interpret the discrepancies analytically in terms of Lamb shift and entanglement with the bath. In Sec.~\ref{sec:gate}, we study the gate error arising from the weak-coupling approximation. We summarize our results in Sec.~\ref{sec:conc}.

\begin{figure}[ht!]
\includegraphics[width=0.96\linewidth]{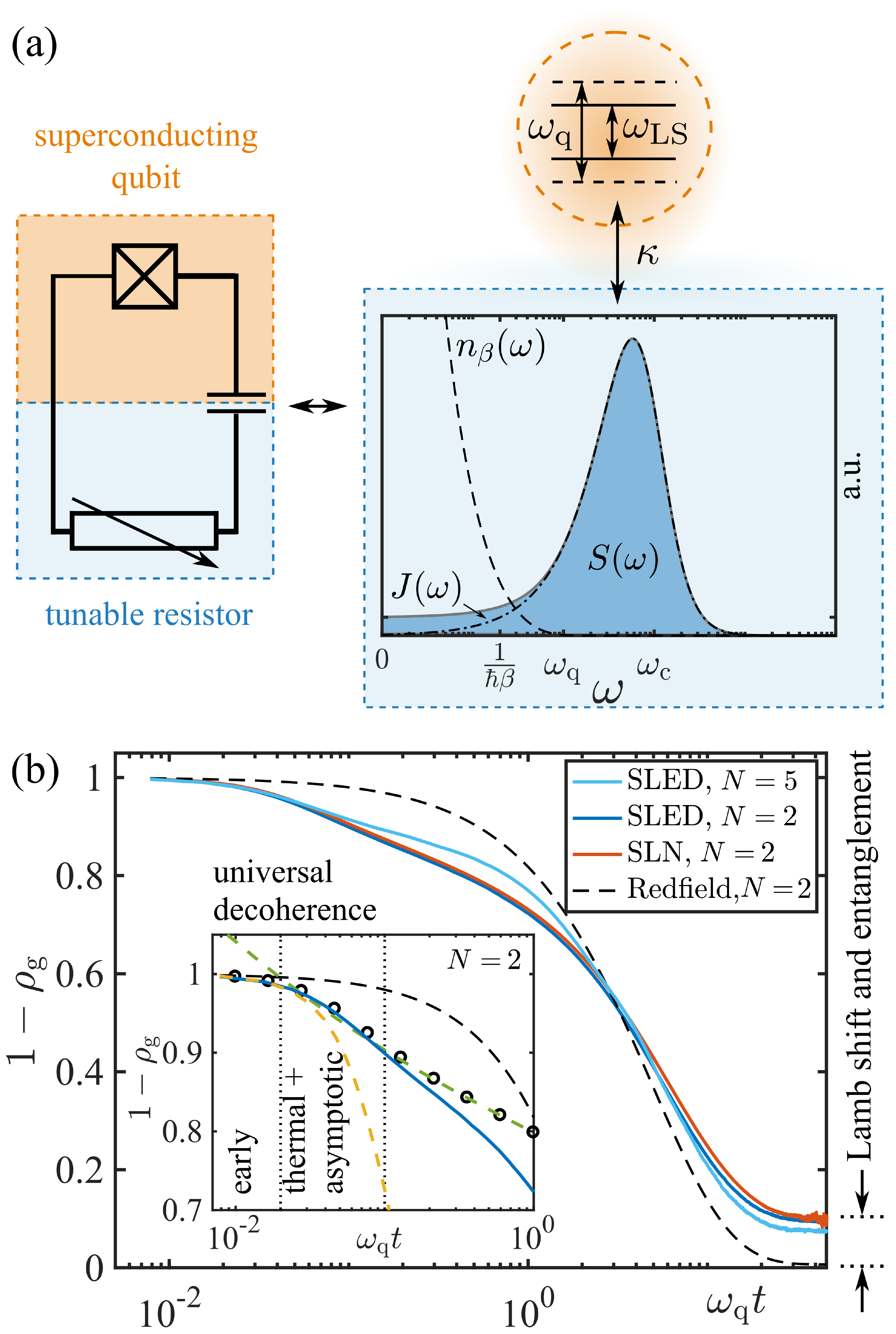}
\caption{(a) \replaced{C}{Schematic illustration of a qubit (dashed orange) interacting with a bath (dashed blue). We show the c}ircuit diagram (left) of a superconducting qubit coupled to a tunable resistor\deleted{ which may be realized using a quantum circuit refrigerator or a tunable heat sink}, together with the corresponding spectral densities (right). The bare qubit angular frequency is denoted by  $\omega_{\rm q}$ and the Lamb-shifted quantity by $\omega_{\rm LS}$. The dynamic properties of the bath are characterized by the power spectrum $S(\omega)$ of the bath fluctuations, and the related mode spectral density $J(\omega)$ and the bosonic occupation $n_\beta(\omega)$. \deleted{The interaction strength between the qubit and the bath is parameterized through the zero-temperature Born--Markov decay rate $\kappa$.}
(b) \replaced{Initialization error for the decay of a qubit excitation}{Decay dynamics of the qubit excitation probability} demonstrating initial universal decoherence, and the Lamb shift and entanglement at long times for the SLN and SLED methods (solid lines). The Redfield solution (dashed line) fails to capture these effects. \added{We show data for ideal ($N=2$) and transmon ($N=5$) qubits.}
Inset: The \replaced{early}{short-time} behavior of the \replaced{initialization error}{excited-state probability}. We show the full universal decoherence (black circles)\replaced{ (see Appendix~\ref{app:unidec})}{~\cite{[{See Supplemental material at xxx}] [{for additional information.}] SI}}, and the \replaced{early}{short}-time approximation $f(t) \approx \frac{1}{2}\, \omega_{\rm c}^2t^2$ (dashed yellow) given in the text. 
The \replaced{green}{pink} dashed line denotes the combined effect of the thermal and asymptotic results in Eq.~(\ref{eq:asymptotic}). The vertical lines at $\omega_{\rm q} t=0.1$ and at $\omega_{\rm c} t =1$ define the regions of validity for the indicated approximations. 
Here \added{$\alpha_{\rm r}=-0.04$,} $\hbar\beta\omega_{\rm q} = 5$, $\kappa/\omega_{\rm q} = 0.2$, and $\omega_{\rm c}/\omega_{\rm q}=50$.}\label{fig:1}
\end{figure}

\section{\replaced{Tunable environment for qubit initialization}{Qubit reset dynamics in tunable environments}}\label{sec:system}
As a generic situation for the qubit reset through a dissipative environment we consider, as shown in Fig.~\ref{fig:1}(a), a superconducting qubit with bare angular frequency $\omega_{\rm q}$ capacitively coupled to a tunable resistor at temperature $T$. The latter is realized using either a quantum-circuit refrigerator\deleted{ (QCR)} or a tunable heat sink\deleted{ (THS)}. A typical power spectral density $S(\omega)$ of such an environment \deleted{subject to the qubit }is also depicted in Fig.~\ref{fig:1}(a) with a maximum around a cutoff frequency $\omega_{\rm c}$ and a zero-frequency limit $\lim_{\omega\to 0}S(\omega)=\kappa/(\hbar\beta\omega_{\rm q})$ where $\beta = 1/(k_{\rm B}T)$ \deleted{is the inverse temperature }and the coupling parameter $\kappa$ is identical to the zero-temperature qubit relaxation rate in the Born--Markov approximation.

These features can be conveniently modeled \replaced{(see Appendix~\ref{app:transmon}) with a}{in terms of a spin-boson-like system with the} dissipative environment bilinearly coupled to the \added{$N$-level transmon} qubit and consisting of an infinite set of harmonic oscillators, i.e.,  
\begin{equation}\label{eq:Ham}
				\hat H = \hbar \sum_{n=0}^{N-1} \omega_n|n\rangle\langle n| + \hbar \sum_k \Omega_k \hat a_k^{\dag}\hat a_k + \hbar \hat q\sum_k g_k(\hat a_k^{\dag}+\hat a_k),
				\end{equation}
where \replaced{$\hat q = \sum_{k,m}\langle k|\hat n |m\rangle |k\rangle\langle m|$, $\hat n$ is the Cooper-pair number operator of the transmon\deleted{~\cite{SI}}, $\omega_n$ and $|n\rangle$ are the eigenfrequencies and eigenstates of the transmon}{$\hat \sigma_{\rm x}=|\textrm{g}\rangle\langle\textrm{e}|+|\textrm{e}\rangle\langle\textrm{g}|$ and $\hat \sigma_{\rm z}=|\textrm{g}\rangle\langle\textrm{g}|-|\textrm{e}\rangle\langle\textrm{e}|$ are Pauli operators, $|\textrm{g}\rangle$ and $|\textrm{e}\rangle$ are the qubit ground and excited states}, respectively, and $\hat a_k$ is the annihilation operator of oscillator mode $k$. \added{The transmon comprises a weakly nonlinear resonator with $\omega_{\rm q} = \omega_1-\omega_0$ and relative anharmonicity $\alpha_{\rm r} = (\omega_2-\omega_1)/\omega_{\rm q}-1$. We restrict our discussion to $N$ lowest energy eigenstates. 
In the limit $N=2$ (ideal qubit), 
Eq.~(\ref{eq:Ham}) reduces to the well-known spin-boson model~\cite{leggett1987,weiss1999} \added{(see Appendix~\ref{app:unidec})} which we use in our analytic calculations and in some numerically exact simulations. We find that $N=5$ (transmon qubit) is enough for accurate studies of the low-energy dynamics at low temperatures.} 

Within this model, the 
power spectrum is obtained as $S(\omega)=J(\omega)[n_{\beta}(\omega)+1]$ with the Bose occupation of the bath modes $n_{\beta}(\omega)=1/[\exp(\hbar\beta\omega)-1]$ 
and the mode spectral density $J(\omega) = 2\pi\sum_k g_k^2\delta(\omega-\omega_k)$, which becomes a smooth function in the limit of a large reservoir. According to Fig.~\ref{fig:1}(a), a \deleted{second-order }Drude model with $J(\omega)=(\kappa/\omega_{\rm q})\, \omega/[1+(\omega/\omega_\textrm{c})^2]^2$ for the tunable resistor captures the relevant physics. \added{Accordingly, the power spectrum gives rise to a Markovian behavior (independent of frequency) only at high temperatures, while at low $T$ it displays a strong frequency dependence inducing non-Markovian dynamics.} Note that the above definitions imply that the ratio $\kappa/\omega_{\rm q}$ is independent of $\omega_{\rm q}$ for $\omega_{\rm q}\ll \omega_{\rm c}$.

Commonly, the \replaced{quantum}{qubit} dynamics within this setting is described with the reduced density operator $\hat \rho$ the time evolution of which is assumed to follow from weak-coupling Redfield- or Lindblad-type master equations (LEs). 
However, the subtle quantum correlations between a qubit and environment require a more sophisticated theoretical treatment that provides predictions which match the experimentally achievable accuracy. 
Suitable methods, originally developed in a condensed matter context~\cite{weiss1999}, have found use in \deleted{the context of }quantum information~\cite{makhlin2001}. Here, the Feynman--Vernon path-integral formalism~\cite{feynman1963}, underlying these methods, is replaced by equivalent stochastic Liouville--von Neumann equation (SLN)~\cite{stockburger1999a,stockburger2002}, unless analytic results exist \added{(see Appendix~\ref{app:exact})}.

The SLN provides an exact non-perturbative treatment of open quantum systems. It augments the Liouville equation with two noise terms which are matched to the free quantum fluctuations of the bath~\footnote{The symmetric and antisymmetric parts of the two-sided spectral function of the bath are related by the fluctuation-dissipation theorem.}. The physical reduced density operator is obtained by averaging over many realizations of the noise. For a \replaced{high}{large} cutoff frequency $\omega_{\rm c}\gg\omega_\textrm{q}$, the SLN equation can be reduced to involve only a single real-valued noise [stochastic Liouville equation with dissipation (SLED)]~\cite{stockburger1998,stockburger1999a}\deleted{ which is beneficial in certain ranges of parameter space}. \deleted{In particular, The SLN and SLED provide an excellent \replaced{tool}{platform} to accurately explore the qubit reset dynamics as they deliver reliable results also for strong qubit--reservoir interactions at low temperatures and \added{even} for time-dependent parameters. }

\section{\replaced{Decay}{Relaxation} dynamics}\label{sec:relaxation}
In Fig.~\ref{fig:1}\replaced{(b)}{(c)}, we monitor the decay of \replaced{the first excited transmon state}{an excited qubit state} as\deleted{, in the presence of a dissipative environment,} it relaxes at low temperatures towards thermal equilibrium. We observe that SLN and SLED results substantially differ from the predictions of the LE during the entire dynamics. Whereas the relaxation follows an exponential decay according to LE, the exact \deleted{qubit }dynamics exhibits various time domains of peculiar behavior. Note that we use in Fig.~\ref{fig:1}(b) a relatively strong environmental coupling, $\kappa = 0.2\times \omega_{\rm q}$, as realized in recent protocols for engineered environments~\cite{tan2017,Silveri2018,partanen2018}. This is outside the strict applicability of the LE. 

For \replaced{early}{short} times, $t\ll 1/\omega_{\rm q}$, the \added{ideal-}qubit dynamics remains frozen and the qubit is only affected by the high-frequency reservoir modes~\cite{braun2001}. As a consequence, \replaced{the initialization error of the qubit}{an excited qubit state} decays as \replaced{$1-\rho_{\rm g}(t) $}{$\rho_{\rm e}(t) = \langle \textrm{e}|\hat{\rho}(t)|\textrm{e}\rangle$}$= \{1+\exp[-f(t)\kappa/(\pi\omega_{\rm q})]\}/2$, where \added{$\rho_{\rm g} = \langle 0|\hat{\rho}|0\rangle$, and} both $f(t)$ and $\kappa/(\pi\omega_{\rm q})$ are system independent quantities, determined only by the reservoir \replaced{(see Appendix~\ref{app:unidec})}{~\cite{SI}}. Such decay, referred to as universal decoherence, can more concisely be described as dephasing in the pointer state basis of $\hat{\sigma}_{\rm x}$\added{$=|0\rangle\langle 1|+|1\rangle\langle 0|$}~\cite{zurek2003,braun2001}. This behavior is depicted in the inset of Fig.~\ref{fig:1}(b), where we observe a good agreement between the analytical prediction and the numerically exact solution if $\omega_{\rm q} t\lesssim 0.1$. In particular, explicit expressions for $f(t)$ \deleted{for Drude-type dissipation }can be found \added{for an ideal qubit} in limiting regimes, namely,  $f(t) \approx \frac{1}{2}\, \omega_{\rm c}^2t^2$ for ultrashort times $\omega_{\rm c} t< 1$ and 
\begin{equation}\label{eq:asymptotic}
f(t) =  2\left(\gamma-\frac12 + \ln(\omega_{\rm c}t)+\ln\left\{\frac{\sinh[\pi t/(\hbar \beta)]}{\pi t/(\hbar \beta)}\right\} \right),
\end{equation}
for times with $\min (t,\hbar\beta) \gg 1/\omega_{\rm c}$. \replaced{Here,}{In this expression} $\gamma\approx 0.577$ denotes the Euler constant. These results indicate that the decay of a qubit excitation is superexponential at the time scale set by $\omega_{\rm c}^{-1}$. Later, there is an algebraic decay at low $T$, especially for experimentally relevant cases of qubit initialization with $(\hbar \beta)^{-1}\ll \omega_{\rm q}\ll \omega_{\rm c}$. The superexponential and asymptotic decays found above are shown in Fig.~\ref{fig:1}(b) and they agree well with the exact solution in their regimes of validity. \added{The difference between the exact two- and multi-level dynamics during the early evolution indicates a leakage to transmon states $|n\rangle$ with $n>1$ at low temperature with $\hbar \beta \omega_{\rm q}=5$. This further validates the conclusion that the exact dynamics cannot be reconciled with a
simple detailed-balance rate structure of the LE.}
\deleted{\replaced{We also note}{Note} that within the domain of universal decoherence, \deleted{Eq.~(\ref{eq:asymptotic}) predicts that} $\rho_{\rm e}\rightarrow \frac12$ for sufficiently large $\kappa$.} We emphasize that the phenomenon of universal decoherence is lost in the coarse-graining procedure underlying the derivation of the LE.

\begin{figure*}[ht!]
\centering
\includegraphics[width=\linewidth]{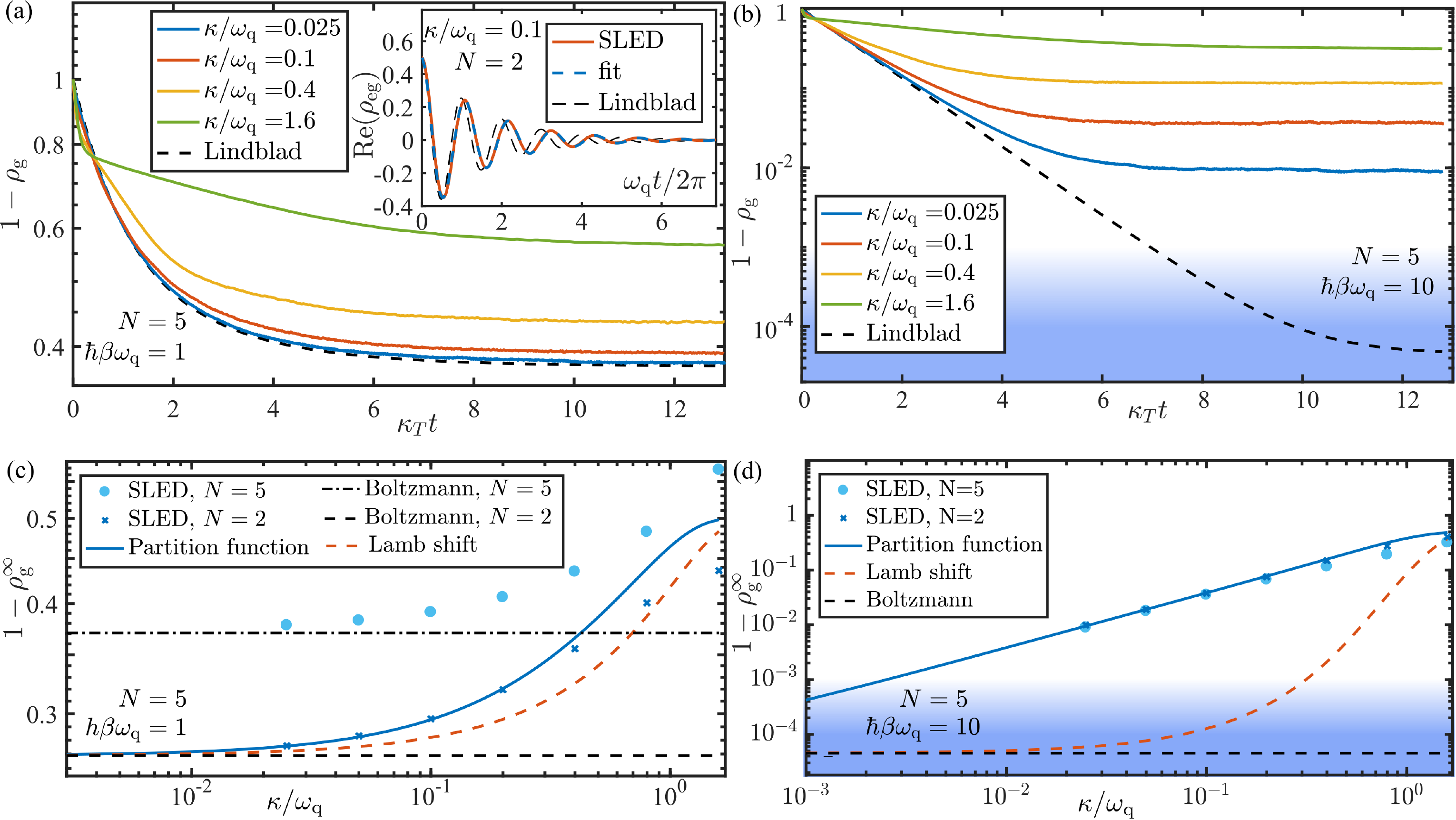}
\caption{(a) \replaced{Initialization error}{Excited-state probability} of a \added{transmon} qubit as a function of time for a high bath temperature, $\hbar\beta\omega_{\rm q}=1$. In all main panels, the \replaced{transmon}{qubit} starts from the \added{first} excited state. Inset: Decaying Larmor precession, i.e., \textrm{Re}($\rho_{\rm eg})$ as a function of time, for a\added{n ideal} qubit initially in the pointer state with $\langle \hat \sigma_{x}\rangle = 1$. We also show an exponentially decaying cosine fit to the SLED data \added{and the corresponding result of the LE}. (b) As (a) but for a low temperature, $\hbar \beta \omega_{\rm q} = 10$, showing the effect of qubit--bath entanglement. (c),(d) \replaced{Initialization error in the steady state}{Steady-state occupation in the excited state} as a function of the bath coupling strength $\kappa$ for (c) the high- and (d) low-temperature data. 
For (c) and (d), the simulation data \added{for $N=5$} (markers) are partly extracted from (a) and (b), respectively, and the analytic partition function result \added{for $N=2$} (solid line) is obtained with Eq.~(\ref{eq:ss}). The color gradient in (b) and (d) indicates the region (blue) feasible for efficient quantum error correction. We have used \added{$\alpha_{\rm r}=-0.04$ and} $\omega_{\rm c}/\omega_{\rm q}= 50$ in all panels. 
In (a) and (b), $\kappa_{\rm T}= \kappa \coth(\hbar \beta \omega_{\rm q}/2)$ is the weak-coupling transition rate from the excited state.}\label{fig:2}
\end{figure*}

\section{Steady-state properties} \label{sec:steady}
Accurate predictions for the qubit steady state are crucial for the fidelity of initialization protocols. Figure~\ref{fig:1}(b) reveals that the bath-coupling strength $\kappa$ may affect the steady-state occupation of the qubit \deleted{quite} substantially. In fact, the ideal Boltzmann distribution at the bare qubit transition frequency $\omega_{\rm q}$ is obtained only in the limit $\kappa\rightarrow 0$. This deviation can be attributed to both the downward Lamb shift of the qubit transition frequency which leads to excess thermal population and the entanglement of the qubit with the \replaced{bath}{residual} degrees of freedom.

In Fig.~\ref{fig:2}, we study in more detail the impact of qubit--reservoir quantum correlations and the role of entanglement as the system approaches the steady state. We \replaced{show}{begin} in Figs.~\ref{fig:2}(a) and~\ref{fig:2}(c) \deleted{with }the dependence of the steady-state probability on $\kappa$ at an elevated temperature, $\hbar\beta\omega_{\rm q}=1$. The \replaced{initialization error}{excited-state occupation} in the steady state \replaced{$1-\rho_{\rm g}^{\infty}$ with $\rho_{\rm g}^{\infty}= \rho_{\rm g}(t\rightarrow \infty)$}{$\rho_{\rm e}^{\infty} = \rho_{\rm e}(t\rightarrow \infty)$} increases with $\kappa$ and the \added{transmon} qubit approaches a fully mixed state already for $\kappa/\omega_{\rm q}>\pi/2$. These numerical findings can further be substantiated \added{in the case of an ideal qubit} by calculating the partition function based on a diagrammatic approach for $\kappa/\omega_{\rm q}\ll 1$ and large cutoff\deleted{~\cite{SI}}, and projecting it on the excited state population \added{(see Appendix~\ref{app:unidec})}. This yields \ $\rho_{\rm e}^\infty = (1-\langle \hat \sigma_{\rm z}\rangle^\infty)/2$, where \added{$\hat \sigma_{\rm z} = |0\rangle\langle 0 |-|1\rangle\langle 1|$ and}~\cite{weiss1999}
\begin{equation}\label{eq:ss}
\langle \hat \sigma_{\rm z}\rangle^\infty = \tanh(\hbar \beta \Omega/2)\frac{\partial \Omega}{\partial \omega_{\rm q}},
\end{equation}
with the renormalized qubit frequency 
\begin{equation}\label{eq:renormalized}
\Omega = \omega_{\rm eff}\left\{1+2 K\left[\textrm{Re}\,\psi\left(i\frac{\hbar \beta \omega_{\rm eff}}{2\pi}\right)-\ln\left(\frac{\hbar \beta \omega_{\rm eff}}{2\pi}\right)\right]\right\}^{1/2}.
\end{equation}
Here, $\psi(x)$ is the digamma function, $\omega_{\rm eff}= G (\omega_{\rm q}/\omega_{\rm c})^{K/(1-K)}\omega_{\rm q}$, $K=\kappa/(\pi\omega_{\rm q})$,
$G=[\Gamma(1-2K)\cos(\pi K)]^{1/[2(1-K)]}$, and
$\Gamma(x)$ is the gamma function. This \deleted{analytic }result agrees well with the exact solution up to $\kappa \approx 0.2\times \omega_{\rm q}$. The inset in Fig.~\ref{fig:2}(a) shows that the decay of the Larmor precession of the pointer state \added{of an ideal qubit} with $\langle \hat \sigma_{x}\rangle =1$ also occurs at this angular frequency $\Omega<\omega_{\rm q}$, which is a signature of the reservoir-induced Lamb shift. More specifically, we find that for $\kappa=0.1\times\omega_{\rm q}$ the renormalized angular frequency $\Omega\approx 0.9\times \omega_{\rm q}$ agrees within 2\% compared to the Lamb-shifted Larmor frequency extracted using a fit for the exact result in the inset. This frequency renormalization reduces to the usual Lamb shift given in the literature only in the ultraweak-coupling limit~\cite{carmichael1999} \added{(see Appendix~\ref{app:unidec})}. \added{We also find that the inclusion of transmon states $|n\rangle$ with $n>1$ renders our system nearly harmonic and, consequently, leads to a suppression of the Lamb shift at high cutoff frequencies (data not shown\added{, see Appendix~\ref{app:unidec}})\deleted{~\cite{SI}}.}

Although one might expect that, at least for very small $\kappa$, the impact of the reservoir on the qubit can be captured solely by a renormalized frequency and the Markovian decay rate, this not the case. Namely, \added{taking} the zero-temperature limit of Eq.~(\ref{eq:ss})\replaced{, we obtain}{reads}
\begin{equation}\label{eq:lowTss}
\rho_{\rm e}^\infty \approx \frac{\kappa}{2\pi\omega_{\rm q}}[-1-\gamma+\ln(\omega_{\rm c}/\omega_{\rm q})]\, ,
\end{equation} 
demonstrating the leading-order correction to the equilibrium state \added{of an ideal qubit} originating from the system--reservoir entanglement \replaced{(see Appendix~\ref{app:unidec})}{~\cite{SI}}. This is in full agreement with our numerical results at a low temperature, $ \hbar\beta\omega_{\rm q}=10$ [cf.~Figs.~\ref{fig:2}(b) and~\ref{fig:2}(d)\deleted{~\cite{SI}}]. Note that this result may exceed the corresponding exponentially small Boltzmann factor by orders of magnitude even for $\kappa$ small enough to leave the transient dynamics virtually unaffected by dissipation. \added{We also show in Fig.~\ref{fig:2}(d) that the transmon states $|n\rangle$ with $n>1$ do not contribute significantly to the initialization error at low temperatures.}

Our findings provide a powerful tool to estimate experimentally achievable qubit initialization fidelities. For example, \deleted{for an }efficient implementation of the surface code \replaced{requires}{one needs} an error below $10^{-3}$ ~\cite{Fowler_2011} which, according to our \deleted{SLN and SLED }results, can only be reached at sufficiently low temperatures and only with coupling strengths $\kappa/\omega_{\rm q}\lesssim 10^{-3}$. This suggests that there exists a trade-off between the speed \deleted{for reservoir-induced reset }and fidelity in \added{reservoir-induced} qubit initialization. In fact, from the low-temperature relaxation rate for weak coupling $\kappa$, the above restriction determines a minimal reset time $\omega_{\rm q} t_{\rm min}\approx 10^4$. Moreover, with a typical angular frequency of superconducting qubits $\omega_\textrm{q}=2\pi\times 8$~GHz and $\omega_\textrm{c}<2\pi\times 400$~GHz, one may use a quantum-circuit refrigerator to tune to $\kappa=10^{-3}\times\omega_\textrm{q}=1/(20\textrm{ ns})$, and hence to reset the qubit to \replaced{$1-\rho_{\rm g}$}{$\rho_\textrm{e}$}$<5\times 10^{-4}$ in less than 200~ns. This would manifest a clear improvement to the current state-of-the-art experiments~\cite{geerlings2013}. \replaced{The}{Note that, according to our findings, the} fidelity in this example cannot be improved by simply lowering the reservoir temperature.


\section{\replaced{Gate error}{Impact on gate operations}} \label{sec:gate}
\begin{figure}[ht!]
\includegraphics[width=\linewidth]{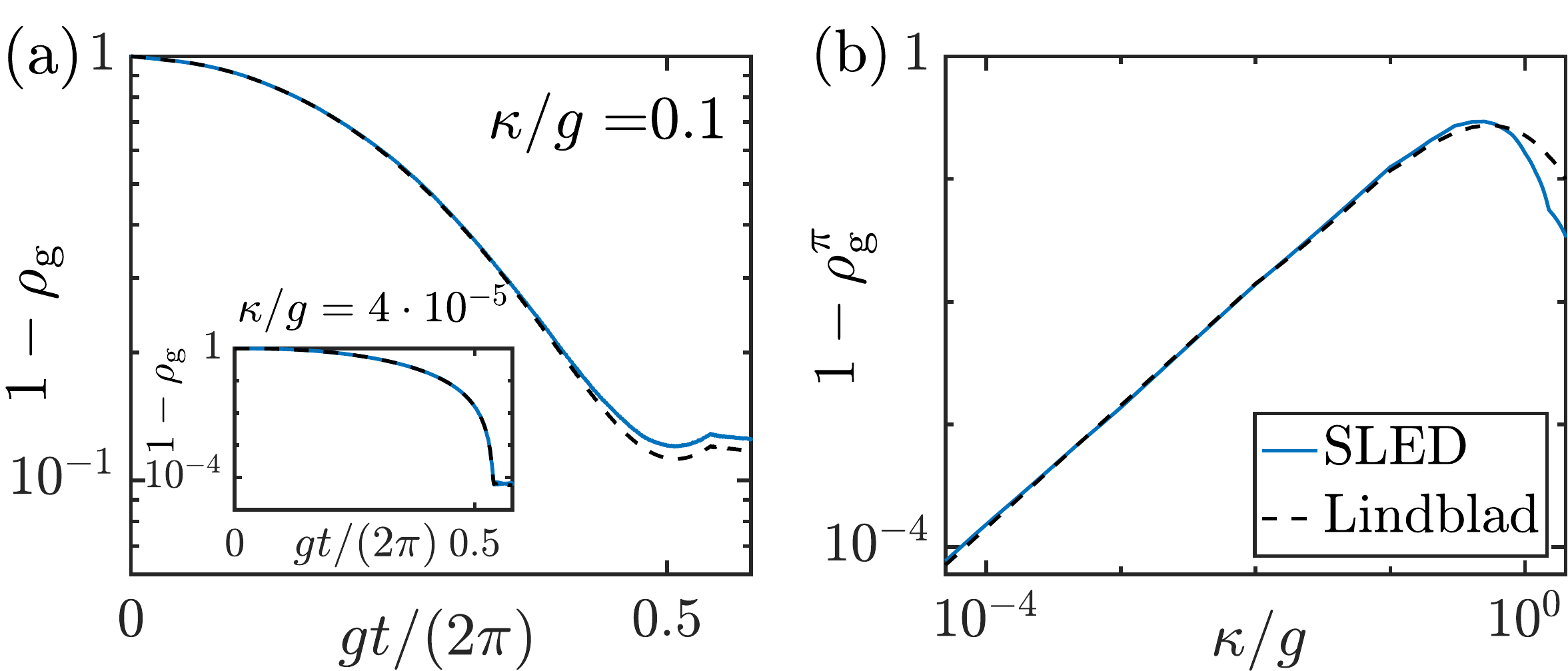}\hfill
\caption{Evolution of the \replaced{gate error}{excited-state occupation} for (a) moderate bath-coupling strength $\kappa$ during the application of a $\pi$ rotation on the \deleted{qubit} excited state \added{of an ideal qubit}. Inset: As (a) but for weak coupling. 
(b) Excited-state population after the gate operation as a function of $\kappa$. We have used $\hbar \beta \omega_{\rm q} = 10$, $g/\omega_{\rm q}=0.0025$, and the rise time of $\pi/(10g)$ for the $\pi$ pulse.}\label{fig:3}
\end{figure}

Given the above subtle qubit--reservoir correlations, the question arises if they may influence also \added{other} qubit \replaced{protocols, such as }{operations other than the reset protocol, for example, }high-fidelity gate operations. We study this \added{for an ideal qubit} in Fig.~\ref{fig:3} for a $\pi$ rotation about $\hat{\sigma}_\textrm{x}$, for various bath coupling strengths compared against the Rabi angular frequency $g$ of the gate.
For $\kappa/g = 4\times 10^{-5}$, the \deleted{conventional }LE provides very accurate predictions \replaced{for the gate error}{to describe the fidelity}, whereas for $\kappa/g = 0.1$, the \replaced{error}{fidelity} is slightly affected by the qubit--reservoir quantum correlations. The gate error as a function of the \replaced{bath}{environmental} coupling strength is depicted in Fig.~\ref{fig:3}(b). Clearly, the error increases with $\kappa$ as the impact of the reservoir leads to a mixing of the qubit state during the pulse. This effect is maximized at $\kappa/g =\frac12$ since we consider only the excited state as the initial state, and hence strong dissipation leads to a fast increase of the desired ground-state population. For \replaced{our}{the} range of parameters\deleted{ used here}, the discrepancies between LE and SLED become relevant for $\kappa \sim g$, a value beyond the practical domain for the implementation of high-fidelity quantum gates. 

\section{Conclusions}\label{sec:conc}
We have shown that the impressive progress in fabricating and controlling devices for quantum information processing calls for non-perturbative approaches beyond conventional weak-coupling master equations to reliably predict the impact of qubit--reservoir correlations for dissipative qubit initialization. In steady state, this includes quantification of the Lamb shift and bath entanglement effects, and the consequent trade-off between initialization speed and accuracy. Fortunately, our results indicate that this trade-off does not seem to pose a fundamental problem on the route to scalable quantum computers if taken properly into account in the initialization protocol. 

Furthermore, we have demonstrated that universal decoherence describes early qubit dynamics up to times of the order of the environment correlation time. This phenomenon is challenging to observe in typical Rabi-driven qubits, but may be visible in cases where the qubit Hamiltonian can be quickly controlled in the laboratory frame~\cite{Nakamura}. Finally, we have observed that the exact dynamics and that given by the Lindblad equation yield matching gate fidelities in parameter ranges of actual implementations. 
Our findings \deleted{contain both caveats and suggestions for novel and intriguing uses of system--reservoir quantum correlations that lie beyond weak-coupling models and} may direct the design of a new generation of state-of-the-art experiments.

\acknowledgments We thank Sahar Alipour\added{, Rebecca Schmidt,} and Matti Silveri for useful discussions. This work was financially supported in part by European Research Council under Grant No. 681311 (QUESS) and by Academy of Finland through its QTF Centre of Excellence program Grants No. 312298 and No. 312300. J.A. and J.T.S. thank the German Science Foundation (DFG) through AN336/12-1 and the IQST for financial support. The authors wish to acknowledge CSC -- IT Center for Science, Finland, for generous computational resources. 

\appendix

\section{\added{Transmon}}\label{app:transmon}
\added{
In the main text, we have studied the system--bath correlations of a superconducting transmon qubit~\cite{koch2007}. The transmon can be modeled with the Cooper-pair-box Hamiltonian
\begin{equation}\label{eq:transmonHam}
\hat H_{\rm S} = 4 E_{\rm C} \hat n - E_{\rm J}\cos \hat \varphi,
\end{equation}
where $E_{\rm C}$ and $E_{\rm J}$ are the charging energy of a Cooper pair and the Josephson energy, respectively, and $\hat \varphi$ and  $\hat n= -i\partial/\partial \varphi$ are the superconducting phase and Cooper-pair number operators of the superconducting island. Contrary to a Cooper-pair box, the transmon is operated in the regime $E_{\rm J}\gg E_{\rm C}$ which leads to a suppression of charge noise. The charge noise depends exponentially on $-\sqrt{E_{\rm J}/E_{\rm C}}$. As a consequence, the Hamiltonian in Eq.~(\ref{eq:transmonHam}) reduces to that of a harmonic oscillator with a weak anharmonicity proportional to $\hat \varphi^4$.
In our exact simulations, we have used the numerically solved angular eigenfrequencies $\omega_n$ and the corresponding eigenvectors $|n\rangle$, with non-negative integer $n$, of the Hamiltonian in Eq.~(\ref{eq:transmonHam}). We have used $E_{\rm J}/E_{\rm C} = 100$, resulting in the relative anharmonicity
\begin{equation}
    \alpha_{\rm r} = \frac{\omega_2-\omega_1}{\omega_1-\omega_0}-1\approx -0.04.
\end{equation}
}
\subsection{\added{Coupling with a harmonic bath}}
\added{
We assume that the transmon is bilinearly coupled through the Cooper-pair-number operator $\hat n$ to a dissipative environment consisting of an infinite set of harmonic oscillators. We represent the total Hamiltonian of the transmon--bath system in the eigenbasis of the transmon as
\begin{equation}\label{eq:transmonbathHam}
	\hat H = \hbar \sum_{n=0}^{N-1} \omega_n|n\rangle\langle n| + \hbar \sum_k \Omega_k \hat a_k^{\dag}\hat a_k + \hbar \hat q\sum_k g_k(\hat a_k^{\dag}+\hat a_k),
\end{equation}
where $\hat q = \sum_{k,m}\langle k|\hat n|m\rangle |k\rangle\langle m|$, and $\hat a_k$ are the annihilation operators of the bath oscillators.
}

\section{Analytic \replaced{early}{short-time} dynamics and steady state}\label{app:unidec}

In the main text, we have presented analytic results for the \replaced{early}{short-time} and asymptotic behavior of the reduced density operator $\hat \rho_{\rm S}$ \replaced{in the case of $N=2$, i.e., an ideal}{ of the} qubit, the time evolution of which is determined by the spin-boson Hamiltonian \added{[$N=2$ in Eq.~(\ref{eq:transmonbathHam})]}
\begin{equation}
\hat H = -\frac{\hbar \omega_{\rm q}}{2}\hat \sigma_{\rm z} + \hbar \sum_k \replaced{\Omega_k}{\omega_k}\hat a_k^{\dag}\hat a_k + \hbar\hat \sigma_{\rm x}\sum_k g_k(\hat a_k^{\dag}+\hat a_k),
\end{equation}
where \replaced{$\omega_{\rm q}=\omega_1-\omega_0$}{$\{\hat a_k\}$ are the annihilation operators of the bath oscillators}, and $\hat \sigma_{\rm z} = |\textrm{g}\rangle\langle \textrm{g}|-|\textrm{e}\rangle\langle \textrm{e}|$ and $\hat \sigma_{\rm x} = |\textrm{g}\rangle\langle \textrm{e}|+|\textrm{e}\rangle\langle \textrm{g}|$ are Pauli matrices where $|\textrm{g}\rangle\added{=|0\rangle}$ and $|\textrm{e}\rangle\added{=|1\rangle}$ are the ground and excited states of the qubit.
Here, we give a detailed derivation of these results.

\subsection{\replaced{Early}{Short-time} decoherence}

In Fig.~\ref{fig:1}(b)\deleted{1(b) of the main text}, we observe that the \replaced{early}{short-time} evolution of \added{the initialization error determined by} the excited-state occupation $\rho_{\rm e} = \langle \textrm{e}|\hat \rho_{\rm S}| \textrm{e}\rangle$ displays a rapid non-exponential drop. The drop occurs at a time scale that is shorter than the characteristic time scale of the system, set by $\omega_{\rm q}^{-1}$. In this \replaced{early}{short}-time limit, one can neglect the free evolution of the system and calculate the decoherence analytically using the Hamiltonian
\begin{equation}
\hat H \approx \hat H_{\rm B} + \hat H_{\rm I} = \hbar \sum_k \Omega_k \hat a^{\dag}_k\hat a_k + \hbar \hat \sigma_{\rm x} \sum_k g_k (\hat a_k^{\dag}+ \hat a_k).
\end{equation} 
The calculation of this so-called universal decoherence was first carried out by Braun \textit{et al.}~\cite{braun2001} by employing the phase-space representation of the density operator for the whole derivation. Here, we give an alternative derivation using the operator formalism.

We assume that initially the system and the bath are statistically independent. Accordingly, the initial density operator can be written as
\begin{equation}\label{eq:factor}
\hat \rho(0)=\hat \rho_{\rm S}(0)\otimes \hat \rho_{\rm B}(0).
\end{equation}
We further assume that the bath oscillators are initially in the thermal state which is determined by the inverse temperature $\beta=1/(k_{\rm B}T)$ and can be expressed as
\begin{equation}\label{eq:thermalinit}
\hat \rho_{\rm B}(0) =\bigotimes_k \hat \rho_k^{\rm th} = \bigotimes_k \frac{1}{Z_k}e^{-\hbar \beta \Omega_k \hat a^{\dag}_k\hat a_k},
\end{equation}
where $Z_k=\textrm{Tr}_{\rm B}\exp(-\hbar \Omega_k\hat a^{\dag}_k\hat a_k)$. In the interaction picture, the von Neumann equation can be written as
\begin{equation}
\frac{\textrm{d}\hat\rho(t)}{\textrm{d} t} = -\frac{i}{\hbar}[\hat H_{\rm I}(t),\hat \rho(t)],
\end{equation}
where 
\begin{equation}
\hat H_{\rm I}(t) = \hbar \hat \sigma_{\rm x} \hat Q(t),
\end{equation}
with $\hat Q(t) =\sum_k g_k (\hat a_k^{\dag}e^{i\Omega_k t}+\hat a_ke^{-i\Omega_k t})$. 
This has the formal solution
\begin{equation}\label{eq:formsol}
\hat \rho(t) = \mathcal{T}e^{-\frac{i}{\hbar}\int_0^t \textrm{d}t' H_{\rm I}(t')}\hat \rho(0) \mathcal{T}e^{\frac{i}{\hbar}\int_0^t \textrm{d}t' H_{\rm I}(t')},
\end{equation}
where $\mathcal{T}$ denotes time ordering. 

Here, we study the operator part $\hat \rho_{nm} = \hat I\otimes\langle n|\hat \rho|m\rangle\otimes \hat I$ of the joint density operator in the eigenbasis of operator $\hat \sigma_{\rm x}$ formed by the pointer states $|n\rangle$ which obey $\hat \sigma_{\rm x} |n\rangle = n|n\rangle$. Here, $\hat I$ is an identity operator for the bath and hence $\hat \rho_{nm}$ is a density operator for the bath.
As a consequence, one obtains
\begin{equation}
\hat\rho_{nm} =  \mathcal{T}e^{-i n\int_0^t \textrm{d}t' \hat Q(t')}\rho_{nm}^{\rm S}(0)\hat \rho_{\rm B}(0) \mathcal{T}e^{i m\int_0^t \textrm{d}t' \hat Q(t')}.
\end{equation}
We express
\begin{equation}
\mathcal{T}e^{im\int_0^t \textrm{d}t' \hat Q(t')} = \lim_{N\rightarrow \infty}\prod_{k=N}^0 e^{im \hat Q(k\delta t)\delta t},
\end{equation}
where $\delta t = t/N$. Iteratively applying the Baker--Campbell--Hausdorff formula with $[\hat a_k,\hat a_l^{\dag}]=\delta_{kl}$ and $[\hat a_k,\hat a_l]=[\hat a_k^{\dag},\hat a_l^{\dag}]=0$, we obtain
\begin{widetext}
\begin{eqnarray}
\mathcal{T}e^{im\int_0^t \textrm{d}t' \hat Q(t')} &=& \lim_{N\rightarrow \infty}\exp\left[im\sum_{k=0}^N Q(k\delta t)\delta t-\frac{m^2}{2}\sum_{k=0}^N\sum_{\ell=0}^{k-1}[\hat Q(k\delta t),\hat Q(\ell \delta t)]\delta t^2\right]\\
&=& \exp\left[im\int_0^t \textrm{d}t' \hat Q(t')-\frac{m^2}{2}\int_0^t\textrm{d}t'\int_0^{t'}\textrm{d}t'' [\hat Q(t'),\hat Q(t'')]\right]\\
&=& \exp\left\{im\sum_k \frac{g_k}{\Omega_k}\left[D_k(t)\hat a_k^{\dag}+D_k^*(t)\hat a_k\right]-im^2\sum_k\frac{g_k^2}{\Omega_k^2}[\Omega_k t-\sin(\Omega_k t)]\right\},
\end{eqnarray}
where we have denoted $D_k(t)=\sin(\Omega_k t) + i [1-\cos(\Omega_k t)]$. 

The reduced density operator is obtained by tracing over the bath degrees of freedom as
\begin{eqnarray}
\rho_{nm}^{\rm S}(t) &=& \textrm{Tr}_{\rm B}\left[\mathcal{T}e^{-in\int_0^t \textrm{d}t' \hat Q(t')}\rho_{nm}^{\rm S}(0)\hat \rho_{\rm B}(0) \mathcal{T}e^{i m\int_0^t \textrm{d}t' \hat Q(t')}\right]\\
&=& \textrm{Tr}_{\rm B}\left[e^{-i(n-m)\int_0^t \textrm{d}t' \hat Q(t')}\hat \rho_{\rm B}(0) \right]e^{i(n^2-m^2)\sum_{\ell} \frac{ g_{\ell}^2}{\omega_{\ell}^2}[\omega_{\ell} t-\sin(\omega_{\ell} t)]}\rho_{nm}^{\rm S}(0)\\
&=& \prod_k \textrm{Tr}_{k}\left\{e^{-i(n-m) \frac{g_k}{\Omega_k}\left[ D_k(t)\hat a_k^{\dag} + D_k^*(t)\hat a_k\right]}\hat \rho_{k}^{\rm th}(0) \right\}e^{i\hbar(n^2-m^2)\sum_{\ell} \frac{ c_{\ell}^2}{2 m_{\ell}\omega_{\ell}^3}[\omega_{\ell} t-\sin(\omega_{\ell} t)]}\rho_{nm}^{\rm S}(0),
\end{eqnarray}
\end{widetext}
where on the last line we have used the thermal initial state for the bath, defined in Eq.~(\ref{eq:thermalinit}). Here, the trace can be simplified with the unitary rotation $\hat U=e^{-i\sum_k \alpha_k(t) \hat a_k^{\dag}\hat a_k}$ where $\alpha_k(t)$ are chosen such that the imaginary part of $D_k(t)\hat a_k^{\dag} + D_k^*(t)\hat a_k$ is eliminated. We also note that the thermal-state density operator is diagonal in the eigenbasis of $\hat a_k^{\dag}\hat a_k$ and, thus, unaffected by the rotation. We obtain
\begin{widetext}
\begin{equation}
\hat U^{\dag} \left[D_k(t) \hat a_k^{\dag} + D_k^*(t)\hat a_k\right]\hat U
= D_k(t) e^{i\alpha_k(t)}\hat a_k^{\dag} + D_k^*(t) e^{-i\alpha_k(t)}\hat a_k
= \sqrt{2[1-\cos(\Omega_k t)]}(\hat a_k^{\dag}+\hat a_k),
\end{equation}
where in the last equality, we have set $\alpha_k(t) = -\textrm{arg}[D_k(t)]$. As a consequence of the transformation, we can write the reduced density operator as
\begin{equation}
\rho_{nm}^{\rm S}(t) = \prod_k \textrm{Tr}_{k}\left[e^{-i(n-m) \frac{g_k}{\Omega_k}\sqrt{2[1-\cos(\Omega_k t)]}(\hat a_k^{\dag}+\hat a_k)}\hat \rho_{k}^{\rm th} \right]e^{i(n^2-m^2)\sum_{\ell} \frac{ g_{\ell}^2}{\omega_{\ell}^2}[\omega_{\ell} t-\sin(\omega_{\ell} t)]}\rho_{nm}^{\rm S}(0).
\end{equation}
\end{widetext}

Thus, we need to calculate $\langle e^{i c_k (\hat a_k^{\dag}+\hat a_k)}\rangle$ in the thermal state, where 
\begin{equation}
c_k= -(n-m)\frac{g_k}{\Omega_k}\sqrt{2[1-\cos(\Omega_k t)]}.
\end{equation}
This can be carried out in the phase-space representation, where the Wigner function for the thermal density operator $\hat \rho_{k}^{\rm th}$ reads
\begin{equation}
W_\beta(q_k, p_k)=\frac{1}{2\pi \sqrt{\langle \hat{q}_k^2\rangle \langle \hat{p}_k^2\rangle}} {\rm e}^{-q_k^2/(2\langle \hat{q}_k^2\rangle) -p_k^2/(2\langle \hat{p}_k^2\rangle)},
\end{equation}
with the thermal averages $\langle \hat{q}_k^2\rangle =q_{k, 0}^2 {\rm coth}(\hbar\beta\Omega_k/2)$ and $\langle \hat{p}_k^2\rangle = \langle q_k^2\rangle \hbar^2/(4q_{k, 0}^2)$ and $q_{k, 0}^2$ being the ground-state width of mode $k$.

As a consequence, one arrives with $\hat q_k=q_{k, 0} (a_k^\dagger+a_k)$ at
\begin{equation}
\langle e^{i c_k (\hat a_k^{\dag}+\hat a_k)}\rangle = e^{-\frac{c_k^2}{2}\coth[\hbar \beta \Omega_k/2]}.
\end{equation}
Finally, we can write the elements of the reduced density matrix as
\begin{equation}
\rho_{nm}^{\rm S}(t) = e^{-(n-m)^2 f(t)\kappa/(4\pi \omega_{\rm q})+i(n^2-m^2) \varphi(t)\kappa/(4\pi\omega_{\rm q})}\rho_{nm}^{\rm S}(0),
\end{equation}
where
\begin{widetext}
\begin{eqnarray}
f(t) &=& \frac{4\pi\omega_{\rm q}}{\kappa}\sum_k \frac{g_k^2}{\Omega_k^2}\coth(\hbar \beta \Omega_k/2)[1-\cos(\Omega_k t)]= \frac{2\omega_{\rm q}}{\kappa}\int_0^{\infty} \textrm{d}\omega \frac{J(\omega)}{\omega^2}\coth(\hbar \beta \omega/2)[1-\cos(\omega t)],\\
\varphi(t) &=&\frac{4\pi\omega_{\rm q}}{\kappa}\sum_{k} \frac{g_{k}^2}{ \omega_{k}^2}[\omega_{k} t-\sin(\omega_{k} t)]=\frac{2\omega_{\rm q}}{\kappa}\int_0^{\infty}\textrm{d}\omega \frac{J(\omega)}{\omega^2}[\omega t-\sin(\omega t)],
\end{eqnarray}
\end{widetext}
and we have recalled that
\begin{equation}
J(\omega) = 2\pi\sum_k g_k^2\delta(\omega-\Omega_k) = \frac{\kappa (\omega/\omega_{\rm q})}{\left(1+\frac{\omega^2}{\omega_c^2}\right)^2},
\end{equation}
where the latter equality holds for an ohmic bath with a second-order Drude cutoff at $\omega_{\rm c}$.
We emphasize that these are identical relations to those obtained previously by Braun \textit{et al.}~\citep{braun2001}.

In the main text, we show data for the excited-state occupation of the operator $\hat \sigma_{\rm z}$ given by
\begin{eqnarray}
\rho_{\rm e}^{\rm S}(t) &=& \frac12[1+(\rho_{-+}+\rho_{+-})] \\
&=& \frac12\left\{1+e^{- f(t)\kappa/(\pi \omega_{\rm q})}[\rho_{-+}(0)+\rho_{+-}(0)]\right\}\\
&=& \frac12[1+e^{-f(t)\kappa/(\pi\omega_{\rm q})}],\label{eq:anfastdrop}
\end{eqnarray}
where in the last equality we have assumed that initially $\rho_{\rm e}(0)=1$. We have denoted 
\begin{equation}
|\textrm{e}\rangle = \frac{1}{\sqrt{2}}(|+\rangle + |-\rangle),
\end{equation}
where the pointer states $|\pm\rangle$ are the eigenstates of the $\hat \sigma_{\rm x}$ operator obeying $\hat \sigma_{\rm x}|\pm\rangle = \pm |\pm\rangle$.

\subsubsection{Very\replaced{ early}{-short-time} behavior}

For very \replaced{early}{short} times ($\omega_{\rm c} t < 1$), the argument in the exponential reduces into
\begin{widetext}
\begin{eqnarray}
f(t) = 2\int_0^{\infty} \textrm{d}\omega \frac{1}{\omega\left(1+\frac{\omega^2}{\omega_{\rm c}^2}\right)^2}\coth(\hbar \beta \omega/2)[1-\cos(\omega t)]&\approx&  t^2\int_0^{\infty} \textrm{d}\omega \frac{\omega}{\left(1+\frac{\omega^2}{\omega_{\rm c}^2}\right)^2}\coth(\hbar \beta \omega/2) \\
&\approx & \frac{ \omega_{\rm c}^2t^2}{2},
\end{eqnarray}\end{widetext}
where in the last equality we have also assumed zero temperature. We thus observe that, contrary to the Fermi's golden rule, the \replaced{early}{short-time} dependence of the excited-state occupation is proportional to $t^2$. This implies that for times obeying $\omega_{\rm c} t <1$, we expect faster than exponential decay of the initial excited-state occupation. This feature is demonstrated in Fig.~\ref{fig:1}(b)\deleted{1(c) of the main text}.

\subsubsection{Thermal part}
In general, the function $f(t)$ that determines the \replaced{early}{short-time} decoherence can be expressed as a sum of the zero-temperature and finite-temperature parts as
\begin{equation}
f(t) = \frac{2\omega_{\rm q}}{\kappa}\int_0^{\infty}\textrm{d}\omega \frac{J(\omega)}{\omega^2}[1+2n_{\beta}(\omega)][1-\cos(\omega t)].
\end{equation}
One obtains an analytic solution for the thermal part by noticing that the Drude cutoff can be neglected at the presence of the thermal cutoff. Thus, one obtains for ohmic spectral density that the finite-temperature part of $f(t)$ is given by
\begin{eqnarray}
f_{\beta}(t) &=& 4\int_0^{\infty}\textrm{d}\omega \frac{n_{\beta}(\omega)}{\omega}[1-\cos(\omega t)]\\
&=& 2\ln\left[\frac{\sinh(\pi t/\hbar\beta)}{\pi t/\hbar\beta}\right].\label{eq:thermal}
\end{eqnarray}
This was obtained by first calculating the integral for the time derivative of $f_{\beta}(t)$ and then integrating the result in time. 

\subsubsection{Asymptotic behavior}
We have calculated symbolically using \textsc{maple} the asymptotic behavior of the zero-temperature part. The result can be written as
\begin{equation}\label{eq:asymp}
f_0(t) =2 \left[\gamma-\frac12 + \ln(\omega_{\rm c}t)\right], \ \ \textrm{for } t\rightarrow \infty,
\end{equation}
where $\gamma\approx 0.577\ldots$ is the Euler constant. We emphasize that for the experimentally relevant case with $(\hbar \beta)^{-1} < \omega_{\rm q}\ll \omega_{\rm c}$, the \replaced{early}{short-time} decoherence is determined accurately by the zero-temperature part of the function $f(t)$ because the thermal time scale is longer than that of the system. The sum of Eqs.~(\ref{eq:asymp}) and (\ref{eq:thermal}) is equal to Eq.~(\ref{eq:asymptotic})\deleted{(2) of the main text}.

\subsection{Partition function approach for the steady state}

The qubit occupation in the steady state can be calculated using the canonical partition function
\begin{equation}
Z=\textrm{Tr}\left(e^{-\beta \hat H}\right)
\end{equation}
of the whole qubit-bath system. For clarity, we study the results in terms of the Kondo parameter $K=\kappa/(\pi \omega_{\rm q})$. If the system depends on a parameter $\lambda$, one can write
\begin{equation}
\frac{\partial}{\partial\lambda}\ln Z = \frac{\textrm{Tr} \left(-\beta \frac{\partial \hat H}{\partial \lambda} e^{-\beta \hat H}\right)}{Z},
\end{equation}
and, consequently,
\begin{equation}\label{eq:partexp}
\left\langle \frac{\partial \hat H}{\partial \lambda}\right\rangle = -\frac{1}{\beta}\frac{\partial}{\partial \lambda}\ln Z.
\end{equation}
Due to the coupling, the expectation values of the qubit in the steady state differ from those obtained with the partition function $Z_0=\exp(\hbar \beta \omega_{\rm q} \hat \sigma_{\rm z})$ of the bare qubit. In the following, we calculate these deviations in the regime of weak coupling with $K$ sufficiently smaller than 1, and show that they are caused by the Lamb shift and the entanglement with the bath.

The equilibrium properties of an open quantum system can be described by a reduced partition function~\cite{weiss1999} $Z_q$ with the property
\begin{equation}
\frac{\partial}{\partial \lambda} \ln Z_q = \frac{\partial}{\partial \lambda} \ln Z
\end{equation}
for any parameter $\lambda$ which appears only in the system Hamiltonian ($\partial H_I/\partial \lambda = \partial H_R/\partial \lambda = 0$).
By setting $\lambda = \omega_{\rm q}$ in Eq.~(\ref{eq:partexp}), one obtains
\begin{equation}\label{eq:sxexpdef}
\langle \hat \sigma_{\rm z}\rangle = \frac{2}{\hbar \beta}\frac{\partial}{\partial \omega_{\rm q}}\ln Z_{\rm q},
\end{equation}
where $\hat \sigma_{\rm z}=|\textrm{g}\rangle\langle \textrm{g}|-|\textrm{e}\rangle\langle\textrm{e}|$ is a Pauli operator of the bare ideal qubit.
The reduced partition function of the dressed qubit can be expressed in the weak coupling regime as~\cite{weiss1999}
\begin{equation}\label{eq:part}
Z_{\rm q} = 2\cosh(\hbar \beta \Omega/2),
\end{equation}
where 
\begin{equation}\label{eq:Omega}
\Omega^2 = \omega_{\rm eff}^2\left\{1+2K\left[\textrm{Re}\,\psi\left(\frac{i\hbar\beta\omega_{\rm eff}}{2\pi}\right)-\textrm{ln}\left(\frac{\hbar\beta\omega_{\rm eff}}{2\pi}\right)\right]\right\},
\end{equation}
$\psi(z)$ is the digamma function, and 
\begin{eqnarray}
\omega_{\rm eff} &=& G\left(\frac{\omega_{\rm q}}{\omega_{\rm c}}\right)^{K/(1-K)}\omega_{\rm q },\\
G &=& \left[\Gamma(1-2K)\cos(\pi K)\right]^{1/[2(1-K)]},
\end{eqnarray}
with $\Gamma(x)$ being the gamma function.
The above result is equivalent to Eq.~(\ref{eq:renormalized})\deleted{(4) of the main text} and valid for all values of $\beta$. It has been derived using an exponential cutoff, but we will show later that this assumption leads into minor deviations from the results given by the Drude cutoff used in the numerical simulations.
Using the chain-rule of derivation in Eq.~(\ref{eq:sxexpdef}), we obtain Eq.~(\ref{eq:ss})\deleted{(3) of the main text}:
\begin{equation}\label{eq:expecsx}
\langle \hat \sigma_{\rm z}\rangle 
= \tanh\left(\hbar \beta \Omega/2\right) \frac{\partial \Omega}{\partial \omega_{\rm q}}.
\end{equation}
The expectation value comprises two factors. We demonstrate below that the first factor describes the Lamb shift due to the renormalization of the qubit frequency by the bath. The other factor, $\partial \Omega/\partial \omega_{\rm q}$, of the expectation value is a measure of entanglement between the qubit and the bath, and can be written as $\partial \Omega/\partial \omega_{\rm q} = \partial \Omega/\partial \omega_{\rm eff}(\partial \omega_{\rm eff}/\partial \omega_{\rm q})$ where
\begin{eqnarray}
\frac{\partial \Omega}{\partial \omega_{\rm eff}} &=& \frac{\Omega}{\omega_{\rm eff}}\\
&& - K\frac{\hbar \beta \omega_{\rm eff}^2}{2\pi\Omega}\left[\textrm{Im } \psi'\left(\frac{i\hbar \beta \omega_{\rm eff}}{2\pi}\right) + \frac{2\pi}{\hbar \beta \omega_{\rm eff}}\right],\nonumber\\
\frac{\partial \omega_{\rm eff}}{\partial \omega_{\rm q}} &=& \frac{G^{1-K}}{1-K}\left(\frac{\omega_{\rm eff}}{\omega_{\rm c}}\right)^K.
\end{eqnarray}

\subsubsection{Lamb shift}

If one neglects the factor 
$\partial \Omega/\partial \omega_{\rm q}$ in the expression in Eq.~(\ref{eq:expecsx}), one obtains
\begin{equation}\label{eq:partLamb}
\langle \hat \sigma_{\rm z}\rangle = \frac{2}{\hbar \beta}\frac{\partial}{\partial \Omega} \ln Z_{\rm q}=\tanh(\hbar \beta \Omega/2),
\end{equation}
where the derivative is with respect to the renormalized frequency $\Omega$ instead of the bare frequency $\omega_{\rm q}$ as in Eq.~(\ref{eq:sxexpdef}).  This describes a qubit with only a frequency renormalization $\omega_{\rm q}\rightarrow \Omega$ which is what one would expect for a system experiencing only the Lamb shift and no entanglement with the bath.

In the literature~\cite{carmichael1999}, the Lamb shift is typically derived using the second-order perturbation theory with respect to the couplings $g_k$. As a consequence, the Lamb-shifted transition frequency of the ideal qubit can be written as
\begin{equation}\begin{split}\label{eq:LSpert}
\omega_{\rm LS} =& \omega_{\rm q}\left\{1+\frac{1}{\pi}\mathcal{P}\int_0^{\infty} d\omega \frac{J(\omega)}{\omega_{\rm q}^2-\omega^2}[1+2n_{\beta}(\omega)]\right\} \\
=& \omega_{\rm q}\left\{1-K\replaced{[-\gamma+\ln(\omega_{\rm c}/\omega_{\rm q})]}{[\gamma+\ln(\omega_{\rm q}/\omega_{\rm c})]}\right\},
\end{split}
\end{equation}
where $\mathcal{P}$ stands for principal value. 
In the second equality, we have assumed zero temperature and $\omega_{\rm c}\gg \omega_{\rm q}$, and used the spectral density $J(\omega)=\pi K \omega \exp(-\omega/\omega_{\rm c})$ with an exponential cutoff. For Drude cutoff, the Euler constant $\gamma=0.577\ldots$ is replaced with $\frac12$. On the other hand, one can make a linear expansion in $K$ for the renormalized qubit frequency given in Eq.~(\ref{eq:Omega}). Note that at zero temperature and for $\omega_{\rm c}\gg \omega_{\rm q}$ the expression (\ref{eq:Omega}) reduces to $\omega_{\rm LS}$ {\em only} under the much stricter constraint $K \ln(\omega_c/\omega_{\rm q})\ll 1$. \added{We also emphasize that for a purely harmonic system with natural frequency of $\omega_{\rm q}$, the Lamb shifted transition frequency is given by~\cite{Silveri2018}
\begin{equation}\begin{split}
    \omega_{\rm LS} =& \omega_{\rm q}\left\{1+\frac{1}{\pi}\mathcal{P}\int_0^{\infty} d\omega \frac{J(\omega)\omega_{\rm q}^2}{\omega(\omega_{\rm q}^2-\omega^2)}\right\}\\ 
    =& \omega_{\rm q}\left\{1+K(1-\gamma)\frac{\omega_{\rm q}}{\omega_{\rm c}}\right\}.
\end{split}
\end{equation}
Thus, contrary to the logarithmic divergence of the Lamb shift of a maximally anharmonic system in Eq.~(\ref{eq:LSpert}), the Lamb shift of a harmonic oscillator converges towards zero with increasing $\omega_{\rm c}$. Being only a weakly anharmonic system, we thus expect that the Lamb shift of the transmon is also small compared to that of a two-level system.
This is also supported by our numerical data (not shown) which display a negligible shift of the Larmor frequency for $N=5$ if compared against the data for $N=2$ shown in the inset of Fig.~\ref{fig:2}(a)\deleted{2(a) of the main text}.
}

\begin{figure}[ht!]
\includegraphics[width=\linewidth]{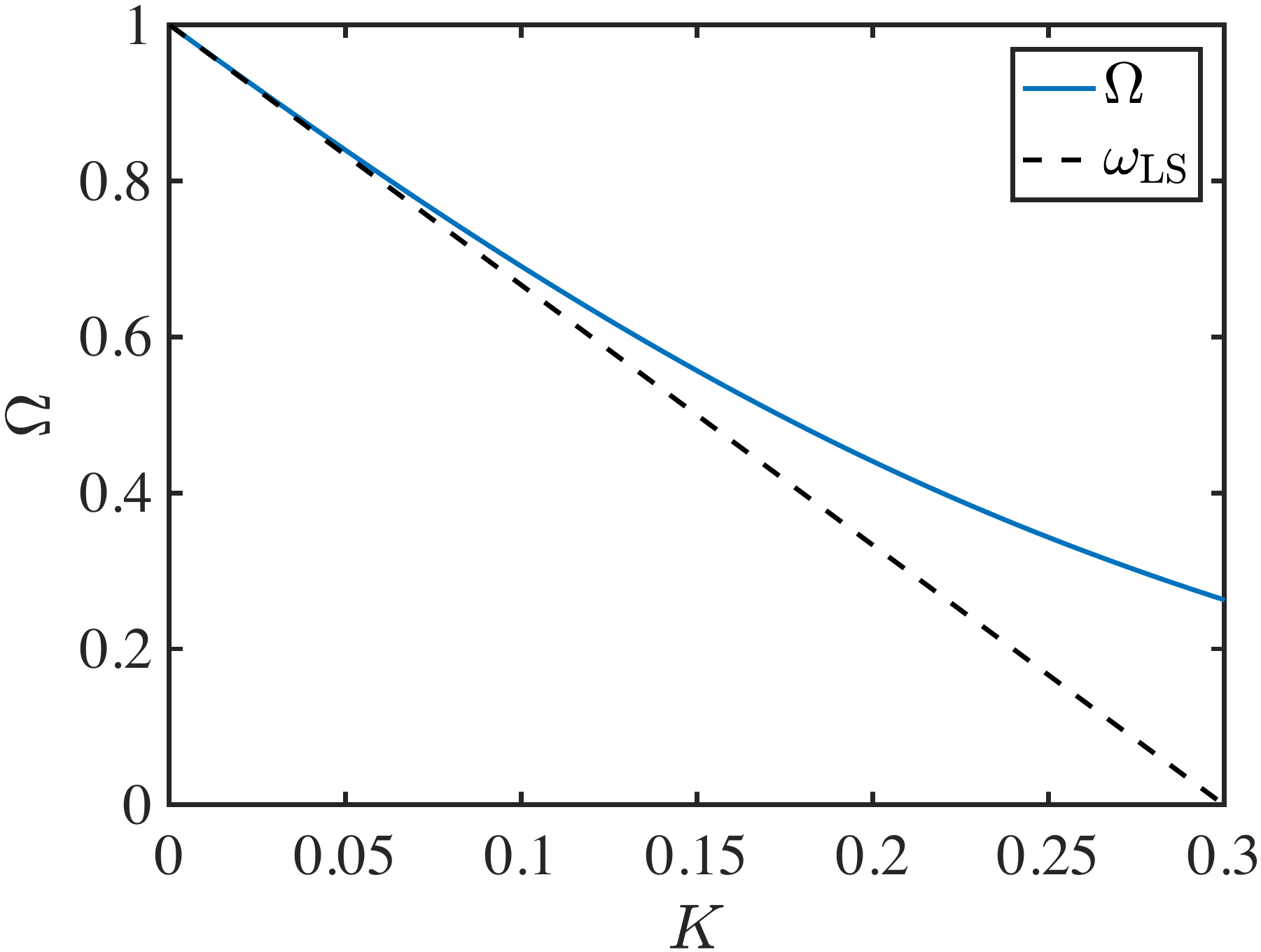}\hfill
\caption{Renormalized qubit frequency $\Omega$ in Eq.~(\ref{eq:Omega}) and the conventional perturbative result $\omega_{\rm LS}$ for the Lamb-shifted qubit frequency defined in Eq.~(\ref{eq:LSpert}) as functions of the Kondo parameter. We have used $\hbar\beta\omega_{\rm q}=5$ and $\omega_{\rm c}/\omega_{\rm q} = 50$.}\label{fig:appfig2}
\end{figure}

In Fig.~\ref{fig:appfig2}, we compare the renormalized frequency $\Omega$ in Eq.~(\ref{eq:Omega}) with the Lamb-shifted qubit frequency $\omega_{\rm LS}$ in Eq.~(\ref{eq:LSpert}). As expected, we observe that the perturbative result follows closely the renormalized frequency in Eq.~(\ref{eq:Omega}) for small values of the coupling constant $K$. At low temperatures with $\beta >5$, the deviations appear for $K\gtrsim 0.1$. The perturbative nature of $\omega_{\rm LS}$ is emphasized by the fact that it decreases without a bound and becomes negative at $K=[\ln(\omega_{\rm c}/\omega_{\rm q})-\gamma]^{-1}\approx 0.3$, where the numerical value has been calculated for the parameters used in Fig.~\ref{fig:appfig2}. On the other hand, the renormalized frequency $\Omega$ approaches zero asymptotically.

However, one cannot obtain Eq.~(\ref{eq:partLamb}) as a limiting case to Eq.~(\ref{eq:expecsx}). This would require $\partial \Omega/\partial \omega_{\rm eff}\rightarrow 1$ and $\partial \omega_{\rm eff}/\partial \omega_{\rm q}\rightarrow 1$. These limits can be reached only at zero temperature and zero $K$, i.e., when the bath can be neglected altogether. Therefore, the steady-state occupation of a qubit is never given by the Boltzmann distribution for the renormalized qubit frequency as the entanglement with the bath generates a notable correction for all $\beta$ and $K$. Especially, when the temperature is zero, the excited-state occupation of the qubit in the steady state is given solely by the entanglement with the bath, as we will show in the following sections.

\subsubsection{Comparison with numerically obtained Larmor frequency}

\begin{figure}[ht!]
\includegraphics[width=\linewidth]{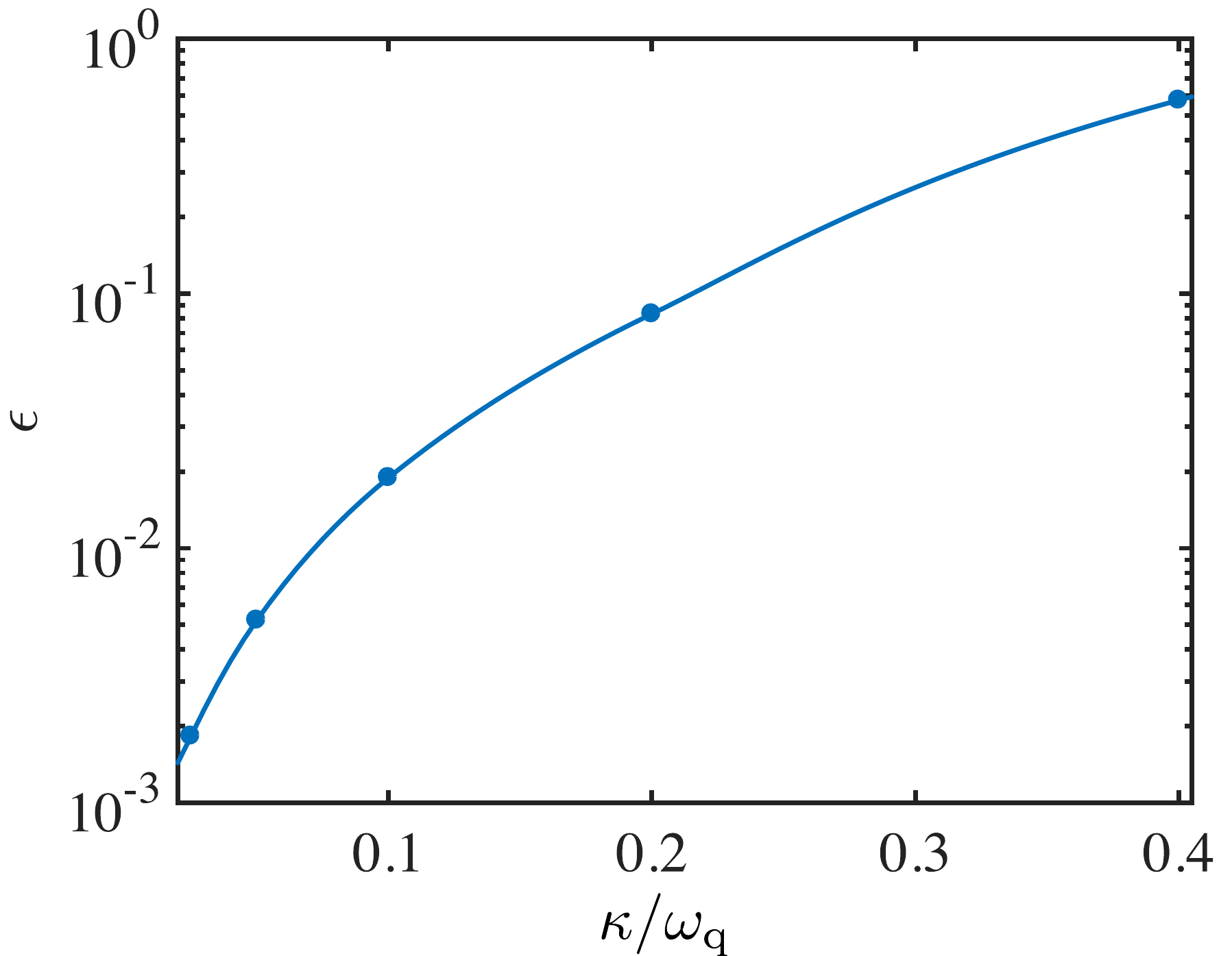}\hfill
\caption{Relative error of the renormalized frequency $\Omega$ in Eq.~(\ref{eq:Omega}) with respect to the numerically obtained Larmor frequency. The Larmor frequencies are obtained by using a pointer state with $\langle \hat \sigma_{\rm x}\rangle = 1$ as the initial state of the SLED simulation. The resulting decaying Larmor oscillations are fitted to a decaying cosine function, similar to the inset of Fig.~\ref{fig:2}(a)\deleted{2(a) in the main text}. We have calculated the relative error numerically for five values of $\kappa$ (dots). We also show an interpolated curve (solid) that goes through the data points. We have used the parameters $\hbar \beta \omega_{\rm q}=1$ and $\omega_{\rm c}/\omega_{\rm q}=50$.}\label{fig:appfig1}
\end{figure}

We have compared the renormalized frequency $\Omega$ in Eq.~(\ref{eq:Omega}) with the numerically obtained Larmor frequency $\omega_0$. In Fig.~\ref{fig:appfig1}, we show the relative error 
\begin{equation}
\epsilon = \frac{|\Omega(\kappa)-\omega_{0}(\kappa)|}{\omega_0(\kappa)}.
\end{equation}
The data shows that for $\kappa \lesssim 0.1\times \omega_{\rm q}$ the relative error is below 2\%.
This further justifies the interpretation of $\Omega$ as the renormalized transition frequency of the qubit in the weak-coupling limit. We also note that the data for $\kappa = 0.1\times \omega_{\rm q}$ correspond to the decaying Larmor oscillations shown in the inset of Fig.~\ref{fig:2}(a)\deleted{\,2(a) of the main text}.

\subsubsection{Zero-temperature occupation}

Here, we show that the coupling to the bath gives rise to a nonvanishing excited-state occupation of the ideal qubit in the steady state, even in the zero-temperature limit $\hbar \beta \omega_{\rm eff}\rightarrow \infty$. We assume a weak coupling ($K\ll 1$) and a high cutoff ($\omega_{\rm c}\gg \omega_{\rm q}$). With these approximations, we obtain in Eq.~(\ref{eq:expecsx}) that $\tanh (\hbar \beta \Omega/2) \approx 1$ and $\Omega \approx \omega_{\rm eff}$. Thus,
\begin{eqnarray}
\langle \hat \sigma_{\rm z}\rangle
&\approx& \frac{\partial \omega_{\rm eff}}{\partial \omega_{\rm q}}\\
&\approx& \frac{\sqrt{\Gamma(1-2K)\cos(\pi K)}}{1-K}\left[1+K\,\textrm{ln}\left(\frac{\omega_{\rm eff}}{\omega_{\rm c}}\right)\right] \\
&\approx & [1+(1+\gamma)K]\left\{1+K\,\textrm{ln}\left(\frac{\omega_{\rm eff}}{\omega_{\rm c}}\right)\right\}.
\end{eqnarray}
We are interested in the steady-state occupation $\rho_{\rm e}^{\rm \infty}=(1-\langle \hat \sigma_{\rm z}\rangle)/2$ in the excited state $|\textrm{e}\rangle$ of the $\hat \sigma_{\rm z}$ operator. 
We obtain up to the first order in $K$ that
\begin{equation}\label{eq:partent}
\rho_{\rm e}^{\infty} \approx \frac{K}{2}\left[-1-\gamma+\textrm{ln}\left(\frac{\omega_{\rm c}}{\omega_{\rm eff}}\right)\right].
\end{equation}

It should be noted that for $K>0$, this result deviates from the $\rho_{\rm e}^{\infty} = 0$ prediction given by the Boltzmann distribution for the bare qubit at $T=0$. Thus, one can interpret that the non-zero occupation of the excited state cannot be treated as a Lamb shift and, thus, has to be generated by the entanglement between the qubit and the bath.

\subsection{Perturbative treatment of the zero-temperature ground-state entanglement}

We calculate perturbatively the reduced density operator of the qubit in the steady state in the zero-temperature limit. We assume that the system is described with the total Hamiltonian
\begin{equation}
\hat H = -\frac{\hbar  \omega_{\rm q}}{2} \hat \sigma_{\rm z}+ \hbar \sum_k \Omega_k\hat a^{\dag}_k\hat a_k +  \hbar\hat\sigma_{\rm x}\sum_k g_k(\hat a_k^{\dag} + \hat a_k),
\end{equation}
and obeys the Boltzmann distribution in the steady state. Thus at zero temperature, the whole qubit-bath system is in its ground state. Since the Jaynes--Cummings type terms $\propto \hat\sigma_+\hat a_k + \hat \sigma_- \hat a_k^{\dag}$ conserve the occupation number, they do not affect the ground state and are, thus, neglected in the following. 
As a consequence, we can write the total Hamiltonian as
\begin{equation}\label{eq:zeroTHam}
\hat H \approx -\frac{\hbar  \omega_{\rm q}}{2} \hat \sigma_{\rm z} + \hbar \sum_k \Omega_k\hat a^{\dag}_k\hat a_k +  \hbar\sum_k g_k(\hat a_k^{\dag}\hat\sigma_+ + \hat a_k\hat \sigma_-),
\end{equation}
where $\sigma_- = |\textrm{g}\rangle\langle \textrm{e}|$ is the annihilation operator of the qubit.
This Hamiltonian can be approximately diagonalized up to the second-order in the couplings $g_k$ with the unitary transformation 
\begin{equation}\label{eq:trans}
\hat U = e^{\sum_k \hat S^{(k)}},
\end{equation}
where $\hat S^{(k)} = [g_k/(\omega_{\rm q}+\Omega_k)](\hat a_k^{\dag}\hat \sigma_+ - \hat a_k \hat \sigma_-)$ diagonalize the interaction terms of the Hamiltonian in Eq.~(\ref{eq:zeroTHam}).  As a result, we obtain the Hamiltonian
\begin{widetext}
\begin{equation}\label{eq:transHam}
\tilde{\hat H} = \hat U \hat H \hat U^{\dag}
\approx  -\frac{\hbar  \omega_{\rm q}}{2} \hat \sigma_{\rm z}  + \hbar \sum_k \Omega_k\hat a^{\dag}_k\hat a_k
+  \hbar\sum_{k,\ell} \frac{g_kg_{\ell}}{\omega_{\rm q}+\Omega_k}\left[\hat \sigma_+\hat \sigma_-(\hat a_k^{\dag}\hat a_{\ell} + \hat a_{\ell}^{\dag}\hat a_k)-\hat \sigma_-\hat \sigma_+(\hat a_{\ell} \hat a_k^{\dag}+ \hat a_k\hat a_{\ell}^{\dag})\right].
\end{equation}
\end{widetext}

We assume that in the transformed frame, the qubit-bath system is in a thermal state at zero temperature, i.e. in the ground state of the Hamiltonian in Eq.~(\ref{eq:transHam}). Since the coupling terms in Eq.~(\ref{eq:transHam}) conserve the occupation number, the ground state can be written as $|\textrm{g},0,0,\ldots\rangle$, where the first quantum number labels the state of the dressed qubit and the rest those of the dressed bath oscillators. 
The corresponding density operator of the qubit-bath system can be written as
\begin{equation}
\tilde{\hat \rho} = |\textrm{g},0,0,\ldots\rangle\langle\textrm{g},0,0,\ldots|.
\end{equation}
The bare qubit occupation can be obtained by transforming the ground-state density operator back to the laboratory frame as
\begin{equation}
\hat \rho = \hat U^{\dag}\tilde{\hat \rho}\hat U \approx \tilde{\hat \rho} + [\tilde{\hat \rho},\hat S] -\hat S \tilde{\hat \rho}\hat S + \frac12 \{\tilde{\hat \rho},\hat S^2\},
\end{equation}
where $\hat S = \sum_k \hat S^{(k)}$, and the second equality holds up to the second order in the coupling constants $\{g_k\}$. The reduced density operator for the qubit is obtained by tracing over the bath, resulting in
\begin{equation}
\hat \rho_{\rm S} = (1-\chi)|\textrm{g}\rangle\langle \textrm{g}| + \chi |\textrm{e}\rangle\langle \textrm{e}|,
\end{equation}
where 
\begin{equation}
\chi = \sum_k\frac{g_k^2}{(\omega_{\rm q}+\Omega_k)^2} = \frac{1}{2\pi}\int_0^{\infty}d\omega \frac{J(\omega)}{(\omega_{\rm q}+\omega)^2},
\end{equation}
is a measure of the entanglement between the qubit and the bath, i.e., hybridization of the reduced density operator of the qubit even in the zero-temperature limit. Above, we have used the definition for the mode spectral density given by
\begin{equation}\label{eq:specdens}
J(\omega) = 2\pi \sum_k g_k^2\delta(\omega-\Omega_k)= \frac{\kappa (\omega/\omega_{\rm q})}{(1+\omega^2/\omega_{\rm c}^2)^2}.
\end{equation}

Finally, let us compare this result with Eq.~(\ref{eq:partent}). For cutoff with $\omega_{\rm c}\gg \omega_{\rm q}$, the excited-state occupation is given by
\begin{equation}
\chi \approx \frac{K}{2}\left[-\frac32 + \ln\left(\frac{\omega_{\rm c}}{\omega_{\rm q}}\right)\right].
\end{equation}
The difference between the constant terms is caused by the fact that Eq.~(\ref{eq:partent}) was derived using an exponential cutoff whereas here we used a Drude cutoff similar to our numerical simulations. Thus, we observe that the $\partial \Omega/\partial \omega_{\rm q}$ factor in Eq.~(\ref{eq:expecsx}) arises due to entanglement. As a consequence, the entanglement is the dominating source of initialization error at low temperatures. 

\section{\added{Effect of intrinsic dissipation}}\label{app:intr}
\added{
In addition to the low-temperature engineered environment described above, the transmon qubit is also coupled to an intrinsic and uncontrollable environment characterized by the zero-temperature dissipation rate $\gamma$ and temperature $T_{\rm i}$. During an initialization protocol, we assume that $\kappa\gg \gamma$. Consequently, the intrinsic environment has a negligible effect on the steady-state occupation of the transmon.}

\added{We show this by studying a harmonic oscillator that is coupled weakly to two thermal baths. We assume weak coupling and, as a consequence, the quantum dynamics of the reduced density operator is governed by the master equation
\begin{equation}\label{eq:lin2bathME}
    \begin{split}
    \frac{\textrm{d}\hat \rho}{\textrm{d}t} =& \frac{\kappa}{2}(N_{\rm ee}+1)[2\hat a\hat \rho \hat a^{\dag}-\hat a^{\dag}\hat a \hat \rho - \hat \rho \hat a^{\dag}\hat a]\\
&+ \frac{\kappa}{2}N_{\rm ee}[2\hat a^{\dag}\hat \rho \hat a-\hat a\hat a^{\dag}\hat \rho - \hat \rho \hat a\hat a^{\dag}]\\ 
    &+ \frac{\gamma}{2}(N_{\rm i}+1)[2\hat a\hat \rho \hat a^{\dag}-\hat a^{\dag}\hat a\hat \rho - \hat \rho \hat a^{\dag}\hat a]\\ 
    &+ \frac{\gamma}{2}N_{\rm i}[2\hat a^{\dag}\hat \rho \hat a-\hat a\hat a^{\dag}\hat \rho - \hat \rho \hat a\hat a^{\dag}],
    \end{split}
\end{equation}
where $N_{\rm ee}$ and $N_{\rm i}$ are the Bose--Einstein occupations of the engineered and intrinsic baths, respectively. We denote the occupation probabilities for the bare oscillator eigenstates $|n\rangle$ with $P_n = \langle n | \hat \rho |n\rangle$. Using the master equation~(\ref{eq:lin2bathME}), one obtains}
\begin{equation}\begin{split}\added{
    \dot{P}_n =&\left[\kappa(N_{\rm ee}+1) + \gamma(N_{\rm i}+1)\right]\left\{ (n+1)P_{n+1}-n P_n\right\} \\
    &+ \left[\kappa N_{\rm ee} + \gamma N_{\rm i}\right]\left\{ nP_{n-1}- (n+1) P_n\right\}.}
\end{split}
\end{equation}
\added{We assume that the steady state ($\dot{P}_n = 0$) is given by the thermal occupation with
\begin{equation}
    P_n = \frac{1}{1+N}\left(\frac{N}{1+N}\right)^n,
\end{equation}
where $N$ is the effective Bose--Einstein occupation for the harmonic oscillator interacting with two independent baths. After straightforward algebra, we obtain
\begin{equation}
    N=\frac{\kappa N_{\rm ee} + \gamma N_{\rm i}}{\kappa+\gamma}.
\end{equation}
Clearly in the limit $\kappa \gg \gamma$ and for a relatively low intrinsic thermal occupation $N_{\rm i}$ of the oscillator, one obtains that $N\approx N_{\rm ee}$. In the main text, we assume that this holds also for relatively large dissipation rate $\kappa \sim 0.1\times\omega_{\rm q}$. As a consequence, we neglect the intrinsic bath in order to clarify our discussions and to emphasize our main message.
}

\section{SLN and SLED methods}\label{app:exact}

If the coupling between the qubit and the bath cannot be treated as a weak perturbation, or the environmental correlation time is long, the typical Born--Markov approximation leading to Redfield and Lindblad master equations becomes inaccurate~\cite{weiss1999}. In such situations, one has to rely on more accurate methods, such as the formally exact Feynman--Vernon path integral formalism. 

\subsection{Stochastic Liouville--von Neumann equation}

For bilinear coupling and the qubit-bath system starting from a factorized initial state given in Eq.~(\ref{eq:factor}), one can show~\cite{stockburger2002} that the path-integral representation for the reduced density operator dynamics can be cast into the form of the so-called stochastic Liouville--von Neumann (SLN) equation
\begin{equation}\label{eq:SLN}
i\hbar \frac{\textrm{d}\hat \rho_{\rm S}}{\textrm{d}t} = [\hat H_{\rm S},\hat \rho_{\rm S}] - \hbar\xi(t)[\hat \sigma_{\rm x},\hat \rho_{\rm S}]-\hbar\nu(t)\{\hat \sigma_{\rm x},\hat \rho_{\rm S}\}. 
\end{equation}
The SLN equation comprises a deterministic coherent part given by the first term on the right-hand side of the equation, followed by the stochastic dissipative part, the dynamical properties of which are set by the complex-valued random variables $\xi$ and $\nu$. The correlations between the qubit and the bath are encoded into the correlations of the noise terms which follow the equations
\begin{eqnarray}
\langle \xi(t)\xi(t')\rangle &=& \textrm{Re} [L(t-t')],\label{eq:zetanoise}\\
\langle \xi(t)\nu(t')\rangle &=&i\Theta(t-t')\textrm{Im}[L(t-t')],\label{eq:crosscorr}\\
\langle \nu(t)\nu(t')\rangle &=& 0\label{eq:nunoise},
\end{eqnarray}
where $\Theta(t)$ is the Heaviside step function and 
\begin{widetext}
\begin{equation}\label{eq:BCSLN}
L(t-t')=\langle \hat\zeta(t)\hat\zeta(t')\rangle = \int_0^{\infty}\frac{\textrm{d}\omega}{2\pi}J(\omega)\{\coth(\hbar\beta\omega/2)\cos[\omega(t-t')]-i\sin[\omega(t-t')]\}
\end{equation}
\end{widetext}
is the autocorrelation function of the bath with $\hat \zeta = \sum_k g_k(\hat a_k^{\dag}+\hat a_k)$.
We note that the correlations $\langle \xi(t)\xi^*(t')\rangle$, $\langle \xi(t)\nu^*(t')\rangle$, and $\langle \nu(t)\nu^*(t')\rangle$ are not fixed by the bath correlation function, and can be thus chosen to optimize the efficiency of the numerical realization.

\subsection{Stochastic Liouville equation with dissipation}

In the case of ohmic dissipation with a high Drude cutoff frequency $\omega_{\rm c}\gg\omega_{\rm q}$, the path integral formalism can be reduced into the form of so-called stochastic Liouville equation with dissipation (SLED)~\cite{stockburger1999,wiedmann2016}
\begin{equation}\begin{split}\label{eq:SLED}
 \frac{\textrm{d}\hat \rho_{\rm S}}{\textrm{d}t} =& -\frac{i}{\hbar}\left([\hat H_{\rm S},\hat \rho_{\rm S}] -\hbar\xi(t)[\hat \sigma_{\rm x},\hat \rho_{\rm S}] \right)\\ 
 &- \frac{\kappa}{2\hbar \beta \omega_{\rm q}}[\hat \sigma_{\rm x},[\hat \sigma_{\rm x},\hat \rho_{\rm S}]] -i\frac{\kappa}{4}[\hat \sigma_{\rm x},\{\hat \sigma_{\rm y},\hat \rho_{\rm S}\}].
 \end{split}
\end{equation}
The above SLED has a stochastic part characterized by a single real-valued noise term $\xi(t)$.  The remaining terms form the deterministic part of the SLED. The autocorrelation function of the noise term is given by the real part of the bath correlation function as
\begin{widetext}
\begin{equation}\label{eq:BCReal}
\langle \xi(t)\xi(t')\rangle = \int_0^{\infty}\frac{\textrm{d}\omega}{2\pi} J(\omega) [\coth(\hbar\beta\omega/2)-2/(\hbar\beta\omega)]\cos[\omega (t-t')].
\end{equation}\end{widetext}
We emphasize that we have treated separately the classical white noise part in the bath correlation function, resulting in convergence of the numerical implementation of the method which is faster than is obtained without such separation.

\section{Implementation of the SLN and SLED methods}

The SLN equation~(\ref{eq:SLN}) and the SLED in Eq.~(\ref{eq:SLED}) can be solved for each noise realization, i.e., sample, using the conventional methods for deterministic differential equations. The samples are generated in a discretized time grid $[0,h,\ldots,(N-1)h]$ with the finite step size $h$, spanning from the initial value at $t=0$ over an interval lasting for several relaxation times $\kappa^{-1}\ll (N-1)h$. 

\subsection{Generation of correlated-noise samples}

The numerical implementation of the correlated-noise samples is the main difference from the corresponding Lindblad algorithm. Here, we describe a numerical scheme for the generation of such samples obeying the correlation functions~(\ref{eq:zetanoise})--(\ref{eq:nunoise}).
We note that these correlation functions arise in the SLN description of the exact path-integral formalism where the bath correlation function $L(t)$ is defined as in Eq.~(\ref{eq:BCSLN}). In the case of the SLED, one only needs to consider Eq.~(\ref{eq:zetanoise}) where the real part of the bath correlation function is defined in Eq.~(\ref{eq:BCReal}). We give here the method of generating the correlated complex-valued noise terms $\xi$ and $\nu$ for the SLN equation, and then discuss in the end how the noise generation is simplified for the case of the SLED.  

We first divide the $\xi$ noise into two parts as $\xi(t)=\xi_{\textrm r}(t) + \xi_{\rm c}(t)$, where $\xi_{\rm r}(t)$ is assumed real and 
\begin{eqnarray}
\langle \xi_{\rm r}(t)\xi_{\rm r}(t')\rangle &=& \textrm{Re} [L(t-t')],\label{eq:corr1}\\
\langle \xi_{\rm c}(t)\nu(t')\rangle &=& i\Theta(t-t')\textrm{Im}[L(t-t')]\nonumber\\
&=&-i\chi_{\rm R}(t-t'),\label{eq:corr2}
\end{eqnarray} 
and $\langle \xi_{\rm c}(t)\xi_{\rm c}(t')\rangle = \langle \xi_{\rm r}(t)\xi_{\rm c}(t')\rangle =\langle \xi_{\rm r}(t)\nu(t')\rangle=0$. 
Moreover, we denote $\xi_{\rm c}(t)=\xi_{\rm c}^{\rm R}(t) + i\xi_{\rm c}^{\rm I}(t)$ and $\nu(t)=\nu^{\rm R}(t) + i\nu^{\rm I}(t)$, where the terms $\xi_{\rm c}^{\rm R,I}$ and $\nu^{\rm R,I}$ are assumed real. These noises can be generated by filtering the independent Gaussian noise samples $x_1(t),x_2(t)$, and $x_3(t)$ with appropriate window functions $W_1(t)$ and $W_2(t)$ as
\begin{widetext}
\begin{eqnarray}
\xi_{\rm r}(t) &=& \int_{-\infty}^{\infty}\textrm{d}t' W_1(t-t')x_1(t') = \frac{1}{2\pi}\int_{-\infty}^{\infty} \textrm{d}\omega W_1(\omega) x_1(\omega) e^{-i\omega t},\label{eq:wind1}\\
\xi_{\rm c}^{\rm R}(t)&=& \int_{-\infty}^{\infty}\textrm{d}t' W_2(t-t')x_2(t') = \frac{1}{2\pi}\int_{-\infty}^{\infty} \textrm{d}\omega W_2(\omega) x_2(\omega) e^{-i\omega t},\\
\xi_{\rm c}^{\rm I}(t) &=& \int_{-\infty}^{\infty}\textrm{d}t' W_2(t-t')x_3(t') = \frac{1}{2\pi}\int_{-\infty}^{\infty} \textrm{d}\omega W_2(\omega) x_3(\omega) e^{-i\omega t},\\
\nu^{\rm R}(t) &=& \int_{-\infty}^{\infty}\textrm{d}t' W_2(t'-t)x_3(t') = -\frac{1}{2\pi}\int_{-\infty}^{\infty} \textrm{d}\omega W_2^*(\omega) x_3(\omega) e^{-i\omega t},\\
\nu^{\rm I}(t) &=& \int_{-\infty}^{\infty}\textrm{d}t' W_2(t'-t)x_2(t') = -\frac{1}{2\pi}\int_{-\infty}^{\infty} \textrm{d}\omega W_2^*(\omega) x_2(\omega) e^{-i\omega t},\label{eq:wind5}
\end{eqnarray}\end{widetext}
where we have defined the Fourier transformation of a function $f(t)$ as
\begin{equation}
f(\omega) = \int_{-\infty}^{\infty}\textrm{d}t f(t)e^{i\omega t}.
\end{equation}
The window functions are determined by the correlation functions in Eqs.~(\ref{eq:corr1}) and~(\ref{eq:corr2}), and can be expressed as
\begin{eqnarray}
W_1(\omega) &=& [L(\omega)-iL_{\rm i}(\omega)]^{1/2},\\
W_2(\omega) &=& \left[\frac12 \chi_{\rm R}(\omega)\right]^{1/2},
\end{eqnarray}
where $L_{\rm i}(\omega)$ is the Fourier transform of Im[$L(t)$].
For odd spectral densities, we can write
\begin{eqnarray}
L(\omega)=S(\omega)&=& J(\omega)[n_{\beta}(\omega)+1],\\
L_{\rm i}(\omega) &=& - i J(\omega)/2,\\
\chi_{\rm R}(\omega) &=& \textrm{Re} [\chi_{\rm R}(\omega)] + i \frac{J(\omega)}{4}\\ 
&=& \frac{J(\omega)}{4}\left[\frac{\omega_{\rm c}}{2\omega}(1-\omega^2/\omega_{\rm c}^2)+i\right],
\end{eqnarray}
where the second equality in the last equation has been written for the ohmic spectral density $J(\omega)$ defined in Eq.~(\ref{eq:specdens}).

Each noise term can be generated numerically by the following protocol:
\begin{itemize}
\item[1.] Produce an array of $N$ independent Gaussian variables $\{x(t_{\ell})\}$ corresponding to the grid points $t_{\ell} = \ell h$ with $\ell = 0,\ldots, N-1$. 
\item[2.] Use fast Fourier transformation on $\{x(t_{\ell})\}$ to obtain $\{x(\Omega_k)\}$ where $\{\Omega_k = k2\pi/(Nh)\}$ define a grid in the frequency space.
\item[3.] Take the inverse Fourier transformation of $W(\Omega_k)x(\Omega_k)$ to obtain the discretized samples of Eqs.~(\ref{eq:wind1})--(\ref{eq:wind5}).
\end{itemize}
In the case of SLED, one needs to do this procedure only once, for $\xi_{\rm r}$, whereas for the SLN equation all five real-valued random variables $\xi_{\rm r},\xi_{\rm c}^{\rm R,I},$ and $\nu^{\rm R,I}$ are needed.

\subsection{Details about the numerical implementation}

The individual solutions for a given sample do not have a physical interpretation but, nevertheless, the density operator can be obtained by taking an average over the solutions obtained with different samples. We have solved the SLN/SLED equations by representing the density operator using a vector notation which allows writing the stochastic equations in the form $\dot{\rho}_{\rm S} = \mathcal{L}\rho_{\rm S}$ where $\mathcal{L}$ is the Liouvillian superoperator of the SLN/SLED equation including both the deterministic and the stochastic parts. For a given noise sample, the ``Liouvillian'' equation is solved deterministically using the Magnus integrator method up to the first order in the time step $h$. 

As both equations can be treated using deterministic methods for each noise sample, the solution for a given sample has the same complexity as the corresponding Lindblad equation. The difference with respect to the conventional Born--Markov master equations arises from the fact that, for a given set of parameters, one has to solve the dynamical equation for many noise samples. The convergence of the averaging procedure depends heavily on the temperature of the bath as the number of needed samples increases rapidly with decreasing temperature.

For the parameters listed in the main text and Table~\ref{tab:params1}, our simulation runs approximately $10^4$ sample points per second on a modern CPU. Because the noise samples are independent, the SLN and SLED methods are easily parallelizable. We have exploited this property at low temperatures where a large number of samples is needed. The parallelization was implemented with supercomputers at CSC -- the Finnish IT Center for Science. The benchmarks for the numerical solution are listed in Table~\ref{tab:params1} for the data relevant for Figs.~\ref{fig:1} and~\ref{fig:2}\deleted{Figures 1 and 2 of the main text}.

\begin{table}[th]
\begin{tabular}{|c|ccccc|}\hline
Figure & $\hbar \beta \omega_{\rm q}$ & $N$ & $N_{\rm S}$ & $\sigma$ & method\\ \hline
1 & 5  & $2^{13}$ \ \ & $2\times 10^7$ \ \ & $1.0\times 10^{-2}$& SLN\\
1 & 5 & $2^{13}$ \ \ & $5\times 10^4$  \ \ & $5.3\times 10^{-4}$& SLED\\
2a & 1 & $2^{12}$ \ \ & $2\times 10^4$ \ \ & $3.0\times 10^{-4}$& SLED\\
2c & 10 & $2^{13}$ \ \ & $2\times 10^5$ \ \ &$7.5\times 10^{-4}$ & SLED\\
\hline
\end{tabular}
\caption{Benchmarks for the numerical solution of the SLN equation and the SLED. We give examples for the data in Figs.~\ref{fig:1} and~\ref{fig:2} of the main text with fixed parameters $\kappa/\omega_{\rm q} = 0.2$ and $\omega_{\rm c}/\omega_{\rm q}=50$. The data have been obtained with the time step $h=2^{-7}\times \omega_{\rm q}^{-1}$. We denote the number of time steps, i.e. the sample length, with $N$ and the number of samples with $N_{\rm S}$. The standard deviation of $\sigma$ in the steady state is calculated in the interval $t\in [9\kappa_{\rm T}^{-1},10\kappa_{\rm T}^{-1}]$ for all data sets. }
\label{tab:params1}
\end{table}

%

\end{document}